\documentclass[12pt]{article}
\usepackage{a4wide}
\usepackage{graphicx}
\usepackage{amssymb}
\usepackage{amsmath}
\usepackage{slashed}
\usepackage{cite}
\newcommand{\be}{\begin{equation}}
\newcommand{\ee}{\end{equation}}
\newcommand{\ba}{\begin{eqnarray}}
\newcommand{\ea}{\end{eqnarray}}

\begin{document}

\begin{titlepage}
\begin{flushright}
\end{flushright}
\vfill
\begin{center}
{\Large\bf Mono-Higgs signature in fermionic dark matter model}
\vfill
{\bf Karim Ghorbani and  Leila Khalkhali }\\[1cm]
{Physics Department, Faculty of Sciences, Arak University, Arak 38156-8-8349, Iran}
\end{center}
\vfill
\begin{abstract}
In light of the Higgs boson discovery, we explore mono-Higgs signature in association with dark matter pair production
at the LHC in a renormalizable model with a fermionic dark matter candidate. 
For two channels with $\gamma\gamma+\text{MET}$ and $b \bar b+\text{MET}$ in the final state
we simulate the SM backgrounds and signal events at $\sqrt{s} = 14$ TeV. We then estimate the LHC sensitivities
for various benchmark points for two integrated luminosities 
${\cal L} = 300~\text{fb}^{-1}$ and ${\cal L} = 3~ \text{ab}^{-1}$. 
We constrain the Yukawa coupling of the dark matter-SM interaction, taking into account bounds from 
mono-Higgs signature, observed dark matter relic density, Higgs physics, perturbativity requirement and 
electroweak measurements. 
Concerning the mono-Higgs search, it turns out that the channel with the largest 
branching ratio, $b \bar b$ channel, provides better sensitivity. There are found 
regions in the parameter space of the model compatible with all the bounds mentioned above
which can be reached in future LHC studies.

\end{abstract}
\vfill
keywords: 
{\bf Dark matter theory, Collider searches, Mono-Higgs
 }
\vfill
{\footnotesize\noindent }

\end{titlepage}

\section{Introduction}
\label{int}
It is well established that dark matter (DM) constitutes about 26\% of the energy-matter content 
of the Universe \cite{Ade:2013zuv,Hinshaw:2012aka}. The problem of dark matter which remains unanswered in the 
standard model (SM) of particle physics can be explained by weak-scale scenarios 
within the freeze-out mechanism in the early Universe \cite{Scherrer:1985zt}. Weakly interacting 
massive particles (WIMPs) are generically well motivated DM candidates in this mechanism, for a review see \cite{Arcadi:2017kky}.
From particle physics vantage point, the main question is what would be the underlaying interactions
between DM and the SM particles. 

Direct detection experiments such as LUX \cite{Akerib:2013tjd} and XENON \cite{Aprile:2012nq} 
are proposed to probe such probable interactions. 
So far in these experiments, there is found no indication of any DM elastic scattering off the target nuclei.
However, the experimental results provide us with bounds on the spin-independent direct 
detection cross section. The upper limits on the elastic scattering cross section may constrain strongly 
the parameter space of theoretical models beyond the SM. It is worth noting that by appealing to some specific DM 
interactions with ordinary matter, the experimental bound can be evaded.   
  
On the other hand, the discovery of the Higgs boson \cite{Aad:2012tfa,Chatrchyan:2012xdj} 
enriched the physics at the scale of electroweak symmetry breaking, i.e., at $\sim \cal{O}$(100) GeV. 
Thus, one intriguing question to ask is whether the nature of dark matter is 
connected in some ways to this weak scale physics. 
The search for the underlying nature behind DM in processes in connection with 
the Higgs production at the LHC is a new avenue as indirect detection of DM.  

One such processes which can happen at the LHC is called mono-Higgs where DM production is accompanied 
by a single Higgs boson in the final state, see mono-Higgs studies within 
both effective field theory approach and simplified models in \cite{Basso:2015aee,No:2015xqa,Petrov:2013nia,Berlin:2014cfa,
Mattelaer:2015haa,Carpenter:2013xra}.
Since dark matter is neutral and interacts weakly with ordinary matter, it will escape 
the detector recoiling against the Higgs, and leaves some amount of missing transverse energy ($\slashed{E}_{T}$ or MET). 
Recent LHC search for dark matter in association with a Higgs can be found in \cite{Aaboud:2017yqz,Aaboud:2017uak}.

Mono-Higgs is among a large class of processes with the production of DM in a collider in association with 
a visible final state X, which is generally dubbed mono-X processes. 
Mono-X searches and studies are carried out for various X, for instance as a light or heavy jet 
in \cite{Khachatryan:2014rra,Aad:2015zva,Aad:2014vea,Beltran:2010ww,Goodman:2010ku,Rajaraman:2011wf,
   Lin:2013sca,Andrea:2011ws,Agram:2013wda,Boucheneb:2014wza}, 
as a Z or W boson in \cite{Aad:2013oja,ATLAS:2014wra,Khachatryan:2014tva,Aad:2014vka,Bell:2012rg,Bai:2012xg,
                       Alves:2015dya,Haisch:2016usn,Bell:2015sza,Bell:2015rdw}, 
and as a photon in \cite{Khachatryan:2014rwa,Aad:2014tda,Fox:2011pm,Abdallah:2015uba}. 
Along the same lines, works with emphasis on pseudoscalar mediator can be found in 
\cite{Buckley:2014fba,Buckley:2015ctj,Buchmueller:2015eea,Berlin:2015wwa,Kozaczuk:2015bea}.

In this work we explore a renormalizable model with fermionic DM candidate which 
has a particular interaction with the SM particles. 
The fermionic DM is connected directly to a real pseudoscalar singlet $\phi$ 
through the operator $\phi \bar{\chi}\gamma^{5}\chi$. Since we would like to stick 
to a renormalizable model, gauge invariance allows only for interaction between 
the pseudoscalar singlet and the SM Higgs such as a dimension-4 operator $\phi^2 H^{\dagger}H$. 

This model has a particular characteristic where its DM candidate can escape not only the current 
direct detection experiments but also the near future experiments like XENON1T. The reason is due to 
the fact that DM-nucleon elastic scattering cross section in this model is velocity 
suppressed \cite{Esch:2013rta,LopezHonorez:2012kv,Pospelov:2011yp}.
Since direct detection experiments puts no constraints in this type of models, it is 
deemed interesting to investigate other avenues such as dark matter production in association 
with mono-Higgs at the LHC. We will study mono-Higgs signals in two important channels: 
$\gamma\gamma + \text{MET}$ and $b \bar b + \text{MET}$ in the final states.
The main purpose in this study is to constrain 
the model parameter space at 95\% confidence level (CL) in mono-Higgs searches 
besides constraints coming from observed DM relic density, invisible Higgs decay measurements 
and electroweak precision measurements. 

This article has the following structure. In the next section, 
we introduce a simplified renormalizable dark matter model.
In section \ref{constitutes} we discuss all possible constraints concerning our model. 
In section \ref{mono-higgs-LHC} we will then investigate the LHC sensitivities to mono-Higgs signature  
at the LHC for two integrated luminosities ${\cal L} = 300~\text{fb}^{-1}$ and ${\cal L} = 3~ \text{ab}^{-1}$.
We conclude in section \ref{conclusion}.

\section{Dark matter model} 
\label{model}

We describe in this section a renormalizable dark matter model introducing a new Dirac field ($\chi$) to 
become our DM candidate and a pseudoscalar field ($\phi$) as a mediator, both being gauge singlet under 
$SU(3)_c \times SU(2)_L \times U(1)_Y$ \cite{Freitas:2015hsa,Ghorbani:2014qpa}.
In this model the CP-invariant interaction Lagrangian consists of an interaction term 
connecting the Dirac and the pseudoscalar fields and an interaction term connecting the pseudoscalar 
to the SM-Higgs doublet $H$, as    
\begin{equation}
\label{int-lag}
{\cal L}_{\text{int}}   =  -ig_{\chi} \phi \bar{\chi}\gamma^{5}\chi - \lambda_{1} \phi^2 H^{\dagger}H  \,.
\end{equation}
The SM-Higgs field gets a non-zero vacuum expectation value $v_{h}$ and we therefore parametrize the Higgs field
as 
\begin{equation}
H = \frac{1}{\sqrt{2}} \left( \begin{array}{c}
                                0  \\
                                v_{h}+ h'
                       \end{array} \right)\,,
\end{equation}
where $v_h = 246$ GeV.
The pseudoscalar potential being CP-invariant is introduced by the Lagrangian  
\begin{equation}
 {\cal L_{\phi}}   =  \frac{1}{2} (\partial_{\mu} \phi)^2 - \frac{m^{2}}{2}\phi^2 -\frac{\lambda}{24}\phi^4  \,,
\end{equation}
and the known SM-Higgs potential is given by 
\begin{equation}
V_{H}  =  \mu^{2}_{H} H^{\dagger}H + \lambda_{H} (H^{\dagger}H)^2  \,.
\end{equation}
It is assumed that the pseudoscalar field acquires a non-zero vacuum expectation value, thus 
\begin{equation}
\phi =  v_{\phi} + s\,.
\end{equation}
It is readily seen that the CP symmetry is no longer preserved since $v_{\phi} \ne 0$.
The second term in the interaction Lagrangian, eq.~(\ref{int-lag}), gives rise to a mixing term in the $h'-s$ mass matrix
for a non-zero $v_{\phi}$. The mass matrix is diagonalized by redefining the scalar fields as 
\begin{equation}
h = \sin \beta~s + \cos \beta~ h'\,, 
\nonumber\\
~~~~~~~~\rho = \cos \beta~ s - \sin \beta~  h'\,, 
\end{equation}
where the mixing angle is defined by 
\begin{equation}
\tan{2 \beta} =  \frac{2\lambda_1 v_{\phi} v_{h}} {\lambda_{H} v_{h}^2 - \lambda v_{\phi}^2/6} \,.
\end{equation}
After field redefinition, the interaction Lagrangian for $\rho$ and $h$ fields becomes
\begin{equation}
\begin{split}
{\cal L}_{\text{int}} + {\cal L}_{\phi}  =  -ig_{\chi} \Big( h \sin \beta  + \rho \cos \beta \Big) \bar{\chi}\gamma^{5}\chi   
 - \Big( \cos^2 \beta \sin \beta \lambda v_{\phi} + 6 \cos \beta \sin^2 \beta \lambda_H v_h \\
   -6 \cos \beta \sin^2 \beta \lambda_1 v_h + 2 \cos \beta \lambda_1 v_h + 6 \sin^3 \beta \lambda_1 v_{\phi}
   -4 \sin \beta \lambda_1 v_{\phi} \Big) \rho^2 h  \\
   - \Big(\cos \beta \sin^2 \beta  \lambda v_{\phi}  -6 \cos^2 \beta \sin \beta \lambda_H v_h - 6 \sin^3 \beta \lambda_1 v_h 
   + 4 \sin \beta \lambda_1 v_h  \\ \hspace{-3cm}
   - 6 \cos \beta \sin^2 \beta \lambda_1 v_{\phi} +2 \cos \beta \lambda_1 v_{\phi}  \Big) h^2 \rho + ... \,,  
\end{split}
\end{equation}
where ellipsis indicate terms with higher number of $h$ and $\rho$ fields.
Moreover, quark interaction with $\rho$ and $h$ fields becomes 
\begin{equation}
 {\cal L} =  -\sum_{q} \frac{m_{q}}{v_h} q\bar q~(h \cos \beta - \rho \sin \beta)\,.
\end{equation}
There is an effective Lagrangian as an extension to the SM which gives us contact interaction 
between the Higgs and photons \cite{Ellis:1975ap,Kniehl:1995tn}. 
After field rotation the effective Lagrangian becomes
\begin{equation}
{\cal L}_{\text{eff}} = -\frac{1}{4} g (h \cos \beta - \rho \sin \beta) F_{\mu\nu} F^{\mu\nu} \,, \\
\end{equation}
with 
\begin{equation}
\begin{split}
 g = \frac{e^2}{4\pi^2 v_h} \frac{47}{18} \Big( 1+\frac{66}{235} \tau_w +\frac{228}{1645} \tau_w^2 
       + \frac{696}{8225} \tau_w^3 + \frac{5248}{90475} \tau_w^4 \\
       + \frac{1280}{29939} \tau_w^5 
       + \frac{54528}{1646645} \tau_w^6  -\frac{56}{705} \tau_t  - \frac{32}{987} \tau_t^2  \Big) \,, \\
\end{split}
\end{equation}
where $\tau_w = \frac{m_{h}^2}{4m_{w}^2}$ and $\tau_t = \frac{m_{h}^2}{4m_{t}^2}$.
The effective coupling $g$ is obtained in the SM after integrating out top quark or W boson in the loops. 
The effective Lagrangian above is employed by the event generator to implement the Higgs decay into diphoton.

In this work we take the mixing angle as a free parameter and instead obtain the quartic couplings in terms of 
the mixing angle and physical masses of the scalars,
\begin{equation}
 \label{couplings}
\lambda_{H}  = \frac{m^{2}_{\rho} \sin^2 \beta +m^{2}_{h} \cos^2 \beta }{2v^{2}_{h}}\,,
\nonumber\\
~~\lambda  = \frac{m^{2}_{\rho} \cos^2 \beta +m^{2}_{h} \sin^2 \beta }{v^{2}_{\phi}/3}\,,
\nonumber\\
~~\lambda_{1} = \frac{m^{2}_{\rho}-m^{2}_{h}}{4v_{h} v_{\phi}} \sin 2\beta.
\end{equation}
We will restrict our numerical computations to regions in the parameter space that the stability of the 
total potential is guaranteed by satisfying the 
relations, $\lambda \lambda_{H} > 6 \lambda^{2}_{1}$ (when $\lambda_1 < 0$), $\lambda > 0$ and $\lambda_{H} > 0$. 
Moreover, the model remains perturbative when we choose $|\lambda_{i}| < 4\pi$.

\section{The constraints in the model}
\label{constitutes}
In this section we introduce all the possible constraints on the DM model prior to the ones we will 
find from mono-Higgs searches.

\subsection{Constraints from Higgs physics and Oblique parameters}

When $m_{\chi} < m_{h} /2 $, the SM Higgs in the above mentioned model can decay invisibly 
into a pair of DM with decay width,
\begin{equation}
\label{dark-decay}
\Gamma_{\text{inv}} = \frac{g_{\chi}^2 m_{h} \sin^2\beta}{8\pi} (1-\frac{4m^{2}_{\chi}}{m^{2}_{h}})^{1/2}.
\end{equation}
The CMS analysis \cite{Chatrchyan:2014tja} which presents a combined searches in two channels, one 
for a SM Higgs production via vector boson fusion and another for a Higgs production 
in association with a Z boson, imposes the strongest bound on the branching ratio 
of the invisibly decaying Higgs.
The analysis found $\text{BR}(h \to \text{inv.}) \lesssim 0.58$ at $2\sigma$ level. 
In addition, bounds on invisible decays of the Higgs boson from Higgs production in association with top quarks
finds $\text{BR}(h \to \text{inv.}) \lesssim 0.24$ \cite{Khachatryan:2016whc}.
Applying this latter bound we find an upper limit for the 
combination of the Higgs-DM coupling and the mixing angle, 
\begin{equation}
|g_{\chi} \tan \beta| < \frac{5.05~(\text{MeV})^{1/2}}{(m^2_{h} - 4 m^{2}_{\chi})^{1/4}} \,, 
\end{equation}
where we used the Standard Model prediction for the Higgs total decay 
width, $\Gamma^{\text{SM}}_{\text{h}} = 4$ MeV \cite{Denner:2011mq}.  

It is possible to constrain the single parameter, $\beta$, by applying the recent measurements on the 
Higgs production and its decay into SM final states \cite{Khachatryan:2016vau}. 
The quantity which is measured by CMS and ATLAS is called signal strength, $\mu$, and is defined as 

\begin{equation}
\mu_{i}^{f} = \frac{(\sigma_{i} \times \text{BR}^{f})_{\text{Exp}}}{(\sigma_{i} \times \text{BR}^{f})_{\text{SM}}} \,.
\end{equation}
The SM Higgs production via channel $i$ is denoted by $\sigma_{i}$ and $\text{BR}^{f}$ is the branching ratio of 
Higgs decaying into SM final state $f$. The combined result obtained by a fit 
over all various production and decay channels reads $\mu = 1.09 \pm 0.1$. 
Since in our model there is mixing between the pseudoscalar and the Higgs, $\sigma_{i}$ is scaled by a factor $\cos^2 \beta$. 
If we assume that $\Gamma^{\text{inv}}_{\text{Higgs}} << \Gamma^{\text{SM}}_{\text{Higgs}}$, then $\text{BR}^{f}$ remains
the same as its SM value. Therefore in this work $\mu \sim \cos^2 \beta$.  
The experimental finding will then restrict the mixing angle $\beta$ to 
values smaller than $\sim 0.1$ at $1\sigma$ precision.

If we consider mixing angles smaller than $\sim 0.1$, then constraints from oblique parameters $S$ and $U$ are 
negligible. When we add an additional (pseudo)scalar to the SM, the oblique parameter $T$ will be modified as follows \cite{Barger:2007im},

\begin{eqnarray}
  T^{BSM} = -\Big(\frac{3}{16\pi s_{w}^2}\Big)
   \Big\{ \cos^2 \beta \Big[\frac{1}{c_{w}^2} (\frac{m_{h}^2}{m_{h}^2-m_{Z}^2}) \ln \frac{m_{h}^2}{m_{Z}^2} 
         - (\frac{m_{h}^2}{m_{h}^2-m_{W}^2}) \ln \frac{m_{h}^2}{m_{W}^2}   \Big]
  \nonumber\\&&\hspace{-10cm}
  + \sin^2 \beta \Big[\frac{1}{c_{w}^2} (\frac{m_{\rho}^2}{m_{\rho}^2-m_{Z}^2}) \ln \frac{m_{\rho}^2}{m_{Z}^2} 
         - (\frac{m_{\rho}^2}{m_{\rho}^2-m_{W}^2}) \ln \frac{m_{\rho}^2}{m_{W}^2}   \Big]  
  \Big\} \,,
\end{eqnarray}
in which $c_{w} = \cos \theta_{W}$. For $\beta = 0$, we have $T^{BSM} = T^{SM}$.
The best fit on electroweak measurements 
dictates $\Delta T = 0.01 \pm 0.12$ \cite{Agashe:2014kda}, where $\Delta T  = T^{BSM} - T^{SM}$. 
The mixing angle is constrained insignificantly if we take $\beta$ less than $\sim 0.1$.

\subsection{DM relic density}
Furthermore, the combined results from Planck and WMAP provide us with the DM relic density, 
$0.1172 < \Omega_\text{DM} \text{h}^2 < 0.1226 $. 
This observation will exclude some regions in the model parameter space.   
In order to find the present amount of DM number density and then the DM relic density for 
various models, it requires the numerical solution of the Boltzmann 
equation at the freeze-out condition,
\begin{equation}
\frac{dn}{dt} +3 H n = - \langle \sigma_{\text{ann}}v_{\text{rel}} \rangle [n^{2}-(n^{\text{eq}})^2 ].
\end{equation}
Here $\langle \sigma_{\text{ann}} v_{\text{rel}} \rangle$ is the thermal average of DM annihilation cross sections times 
the relative velocity and $n^{\text{eq}}$ is the total number of $\chi$ particles at equilibrium.

In this work, to obtain the DM relic density we make use of the numerical package microOMEGAs \cite{Belanger:2013oya} which 
exploits the package CalcHEP \cite{Belyaev:2012qa} to compute the relevant annihilation 
cross sections. One may consult \cite{Ghorbani:2014qpa} for some detailed DM phenomenology of the 
model under discussion in the present work.

\section{Mono-Higgs signature at the LHC }
\label{mono-higgs-LHC}
\begin{figure}
\begin{center}
\includegraphics[scale=.75,angle =0]{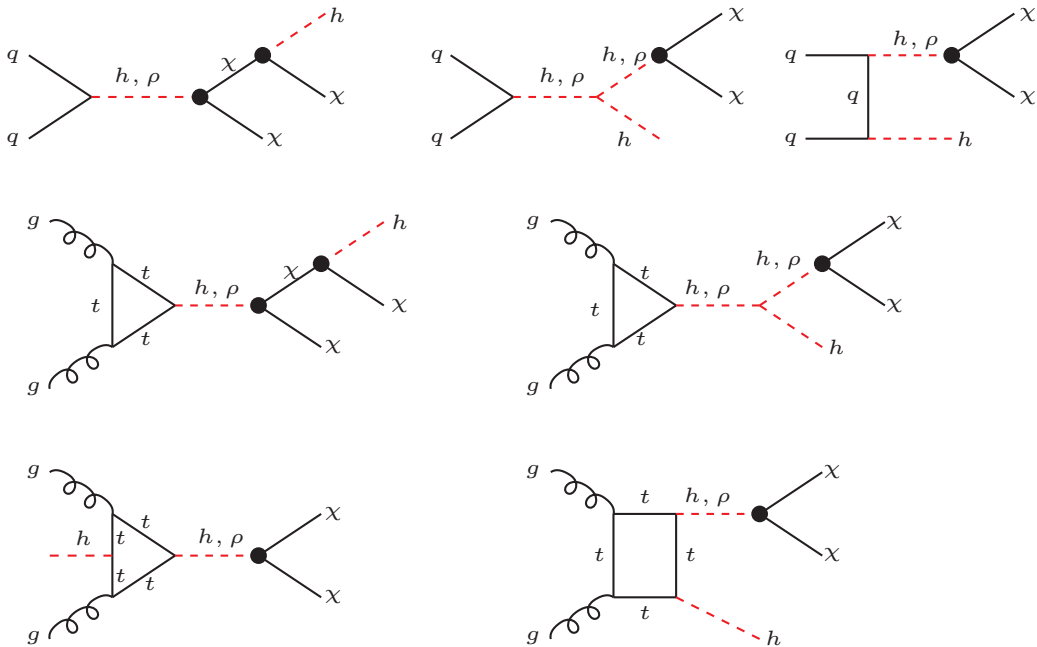}
\end{center}
\caption{Feynman diagrams relevant for the mono-Higgs production at the LHC in the process $p p \to \chi \chi h$.}
\label{diagrams}
\end{figure}

We devote this section to study the LHC sensitivity to mono-Higgs plus MET production at $\sqrt{s} = 14$ TeV 
for two prospective integrated luminosities, ${\cal L} = 300~\text{fb}^{-1}$ and ${\cal L} = 3~ \text{ab}^{-1}$.
We generate our signal and background unweighted events at leading order in MadGraph5aMC@NLO v2.1.2 
\cite{Alwall:2014hca,Hirschi:2015iia} with the CTEQ6L1 PDF \cite{Pumplin:2002vw} which are then passed on to 
PYTHIA6 \cite{Sjostrand:2006za} for patron showering and hadronization. 
For detector simulation we employ Delphes 3 \cite{deFavereau:2013fsa} (to simulate CMS detector) which integrates 
Fastjet \cite{Cacciari:2011ma} with $R = 0.5$ to allow jet reconstruction 
using the anti-$k_{t}$ algorithm \cite{Cacciari:2008gp} as jet clustering algorithm. 
We employ MadAnalysis 5 \cite{Conte:2012fm,Dumont:2014tja} to perform the analysis in this work.

In this study we take the signal production cross section at leading order but the production cross 
sections for various background samples will be corrected by using appropriate $k$-factors 
so as to normalize them to higher order calculations. Therefore, our signal efficiencies 
in this work are somewhat underestimated.    

\begin{figure}
\begin{minipage}{0.5\textwidth}
\includegraphics[width=\textwidth,angle =0]{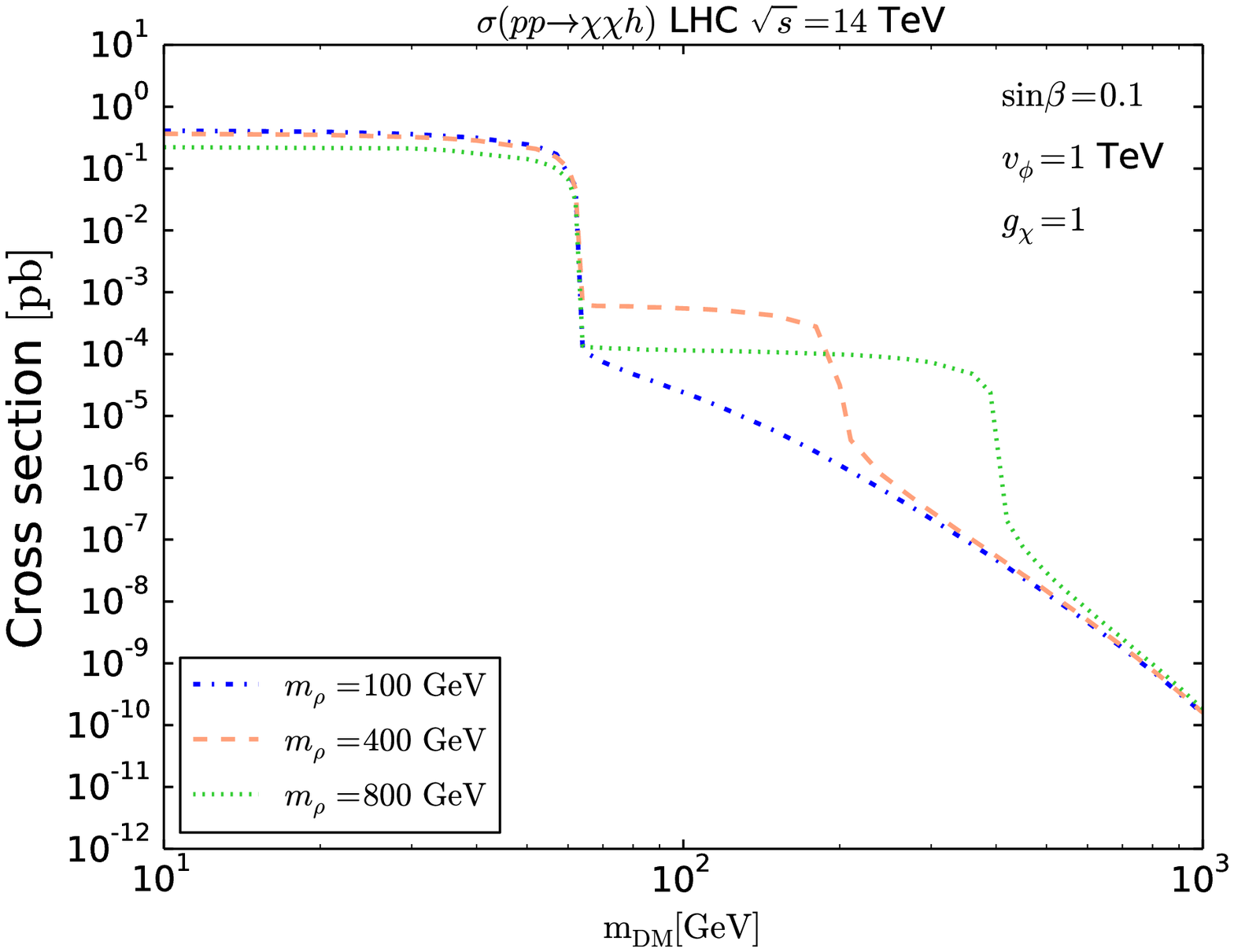}
\end{minipage}
\hspace{.5cm}
\begin{minipage}{0.5\textwidth}
\includegraphics[width=\textwidth,angle =0]{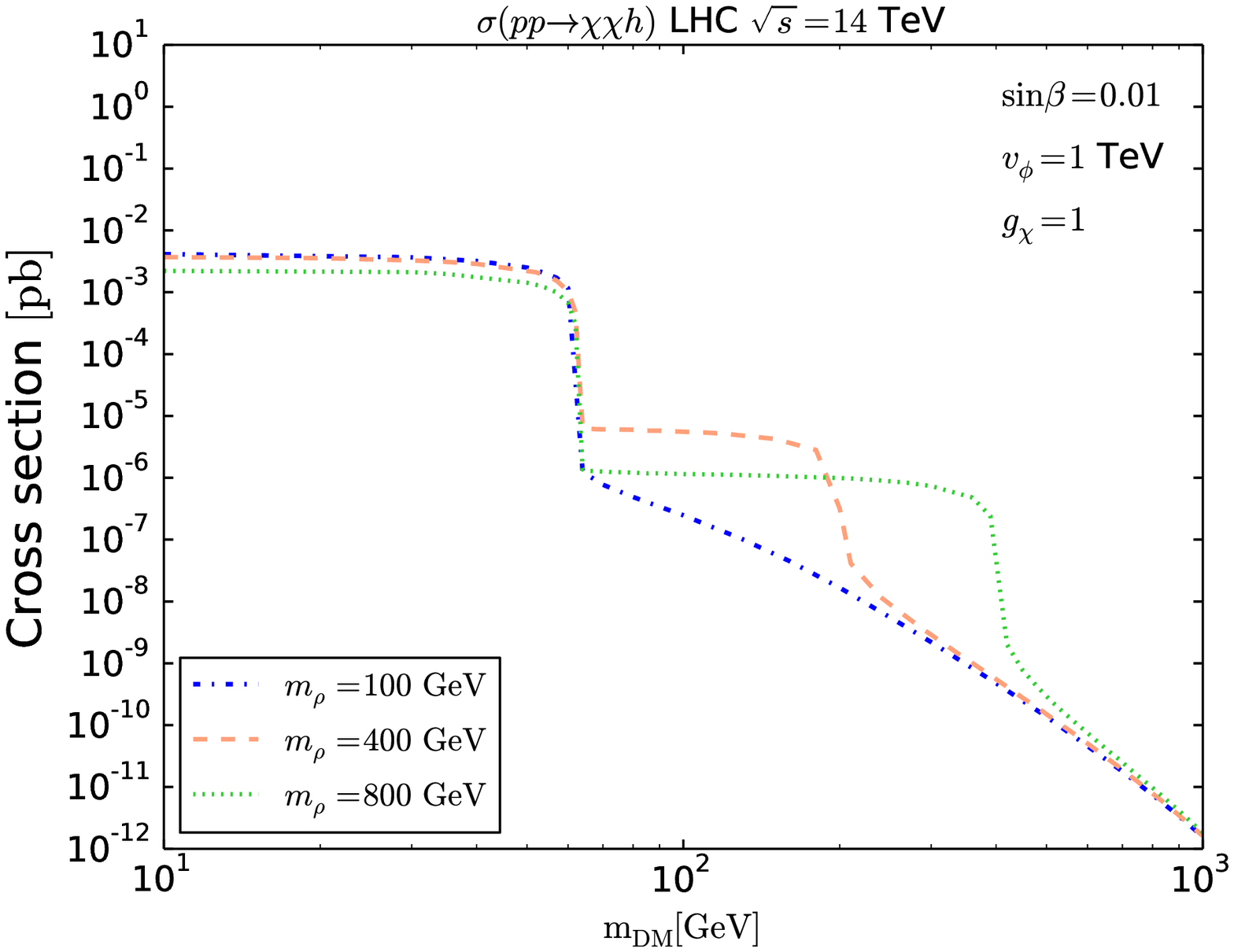}
\end{minipage}
\caption{Mono-Higgs production cross section via the process $p p \to \chi \chi h$ for various benchmark models at 
$\sqrt{s} = 14$ TeV for $v_{\phi} =$ 1 TeV, $\sin \beta = 0.1$ (left) and $\sin \beta = 0.01$ (right).}
\label{sig-production}
\end{figure}

For the signal process with missing transverse energy plus Higgs in the final state, $p p \to \chi \chi h$, we show the relevant 
Feynman diagrams in Fig.~\ref{diagrams}. 
This process is induced through quark or gluon fusions in $pp$ collisions via s-, t- and u-channels. 
In our numerical investigation we find out that diagrams with gluon fusion have dominant contributions
to the total cross section. 
Our numerical results for the signal production cross section for various benchmark 
points at $\sqrt{s} = 14$ TeV are presented as a function of DM mass in 
Fig.~\ref{sig-production} for $\sin \beta = 0.1$ and $\sin \beta = 0.01$. 
The results shown for three mediator masses $m_{\rho} = 100 , 400$ and $800$ GeV, indicate strong 
dependency of the signal production cross section on the mediator mass when $m_{\text{DM}} > m_h /2$. 
In the region with $m_{\text{DM}} < m_h /2$, since the Higgs resonance is accessible it dominates the cross
section. In this region the signal cross sections remain almost steady and then drop off. The same 
reasoning can be applied for the $\rho$ resonance. 
We have checked that the signal cross section does not change significantly by going 
from $v_{\phi} = 0.5$ TeV to $v_{\phi} = 1$ TeV. Moreover, the signal production cross 
section is proportional to $g_{\chi}^2$ based on the Feynman diagrams in Fig.~\ref{diagrams}. 
We choose $v_{\phi} = 1$ TeV, $g_{\chi} = 1$ and $\sin \beta = 0.1$ throughout our computations. 

Our focus in this research is to estimate the LHC sensitivity reach in two 
important Higgs decay channels: $h \to \gamma \gamma$ and $h \to b \bar b$.

\subsection{Two photon channel}
This decay mode is important because we deal with small background events which 
make a clean environment for experimental measurements. 
However, the Higgs decay into two photons has a small branching 
ratio, $\text{Br}(h \to \gamma \gamma) = 2.28 \times 10^{-3}$, and hence the signal production 
rate will be comparatively low. 

\begin{figure}
\begin{minipage}{0.5\textwidth}
\includegraphics[width=\textwidth,angle =0]{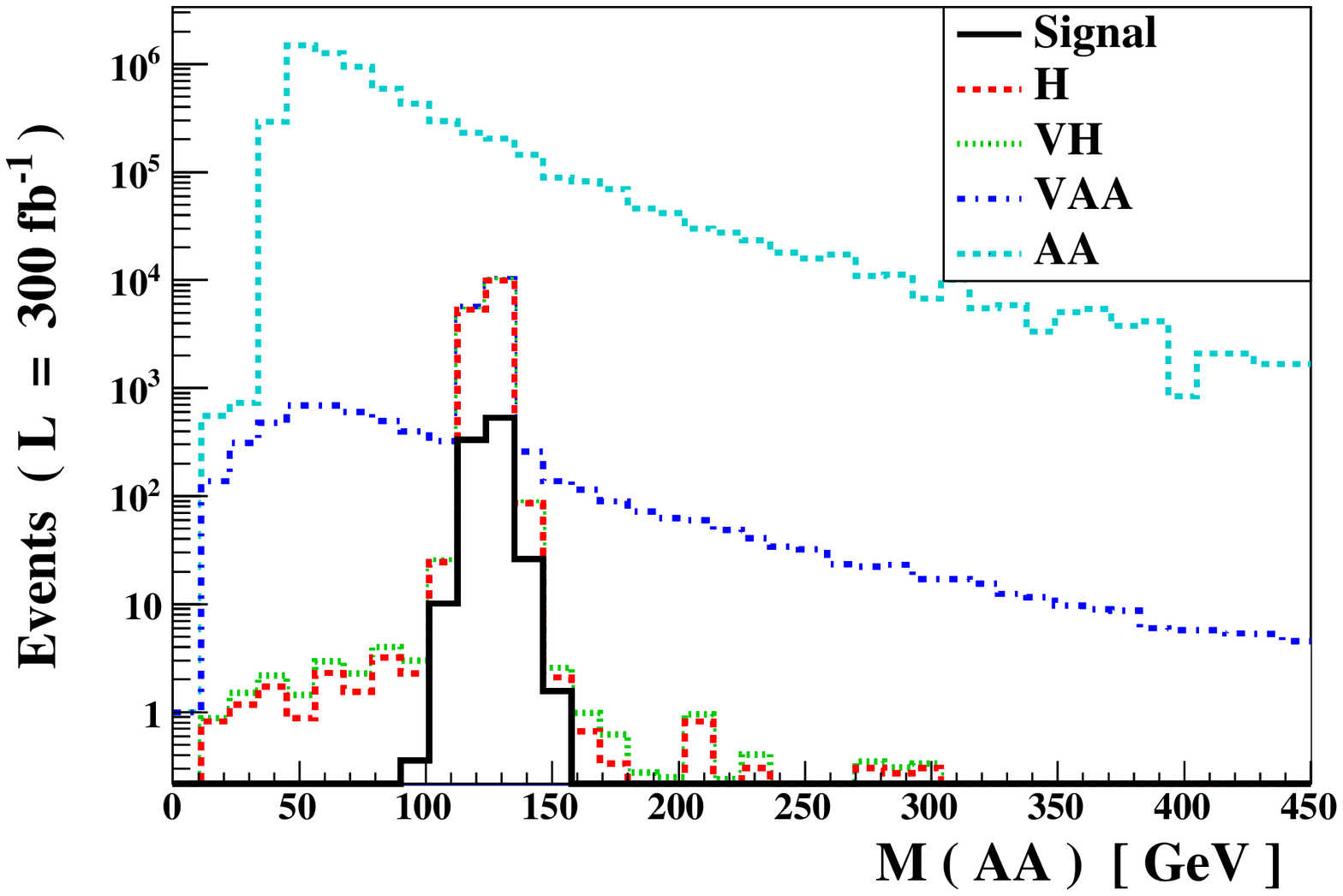}
\end{minipage}
\hspace{.1cm}
\begin{minipage}{0.5\textwidth}
\includegraphics[width=\textwidth,angle =0]{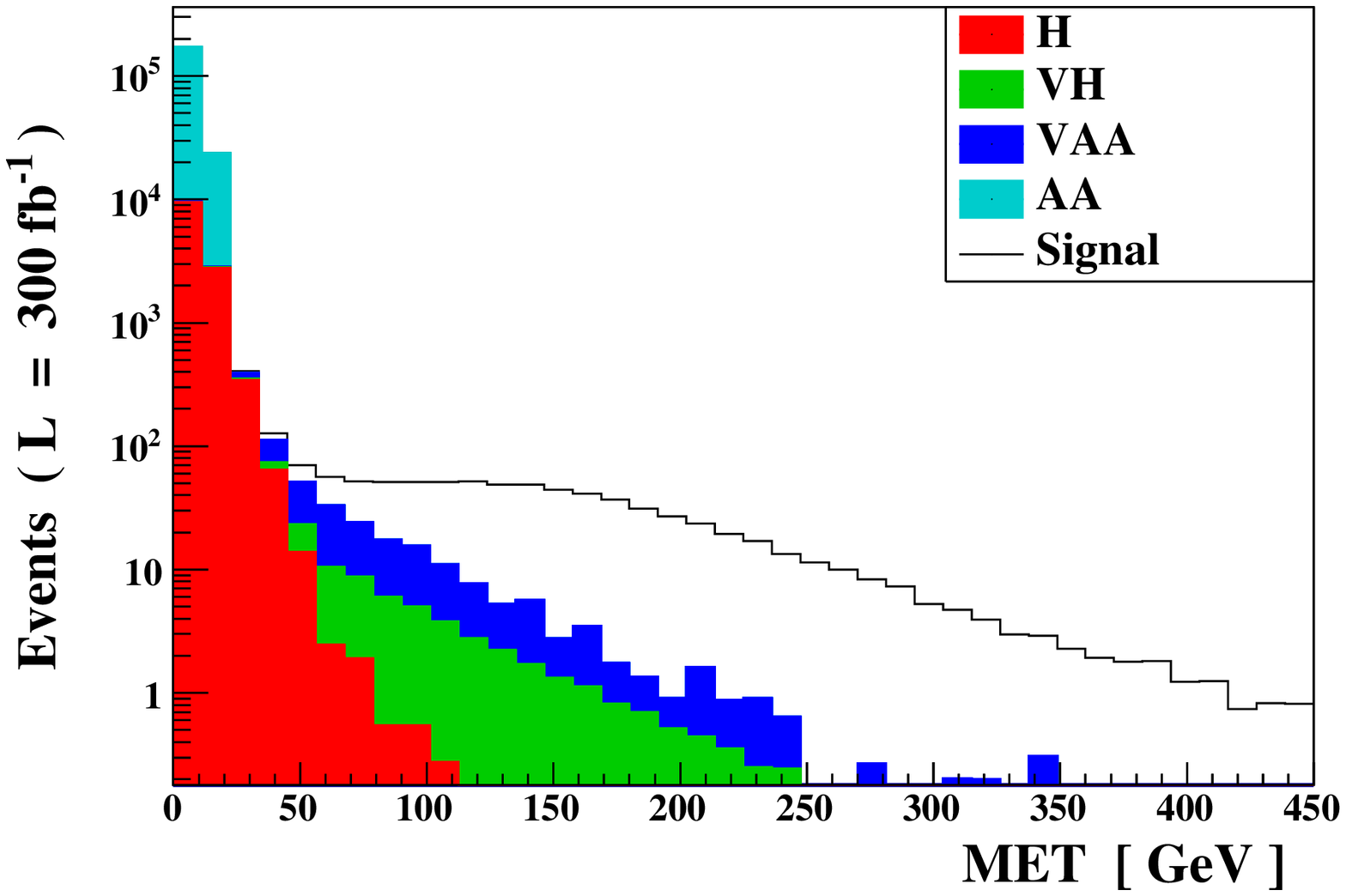}
\end{minipage}
\caption{For $\gamma\gamma+\text{MET}$ channel at $\sqrt{s} = 14$ TeV 
with $\cal{L} = $ $300~\text{fb}^{-1}$, in the left panel 
is the diphoton invariant mass for the signal and background processes  
and in the right panel is distributions of the missing transverse energy for the 
signal and backgrounds after applying the event selections discussed in the text. 
Signal is normalized to a nominal cross section of $5$ fb.}
\label{aa-invariant}
\end{figure}

Important backgrounds to the final state $\gamma\gamma+$MET are as we list:

\begin{enumerate}
   \item $Z h$ with $h \to \gamma\gamma$ and ($Z \to \nu \bar \nu$ or $Z \to \tau^+ \tau^-$ where $\tau$ 
      decay produces some missing energy). 
    This is an irreducible background and is denoted by VH in plots.
   \item $W h$ with $h \to \gamma\gamma$ and $W \to l \nu$, denoted by VH.
   \item $Z\gamma\gamma$ with $Z \to \nu \nu$ or $Z \to \tau^+ \tau^-$ (where $\tau$ decay produces some missing energy), 
      denoted by VAA.
   \item $W\gamma\gamma$ with $W \to l \nu$, denoted by VAA.
   \item $\gamma\gamma$ through Higgs production or via non-resonant production, denoted by H and AA respectively. 
\end{enumerate}
We consider as event selections, the existence of two photons in the final state with transverse momentum $p_{T} > 20$ GeV 
and rapidity $|\mu| < 2.5$ and we veto electrons or muons with $p_{T} > 20$ GeV and $|\mu| < 2.5$. 
In order to improve signal to background efficiency we impose the cut   
$120~\text{GeV} < m_{\gamma\gamma} < 130~\text{GeV}$ for invariant mass of diphoton.  

In Fig.~\ref{aa-invariant} we present the diphoton invariant mass after the event selections are applied 
and also distributions for signal and background events after both the event selections and 
the cut are taken into account. 
The SM backgrounds are estimated using $k$-factors to normalize the leading order (LO) cross sections 
to their values at NLO or beyond.
We use $k = 1.65$ \cite{Campbell:2012ft} for $Z\gamma\gamma$ and the same value for $W\gamma\gamma$, 
$k= 1.3$ \cite{Maltoni:2013sma} for $Zh$ and the same value 
for $Wh$, $k= 1.8$ \cite{Heinemeyer:2013tqa} for Higgs production which includes NNLO QCD and NLO EW 
corrections and $k= 1.6$ \cite{Aad:2011mh} for $\gamma\gamma$ production.

\begin{table}
\centering
\begin{tabular}{p{4.5cm} p{3cm}}
 \hline \hline
    $pp \to \gamma\gamma+ \slashed{E}_{T}$    &    $\slashed{E}_{T} > 150$ GeV \\ \hline \hline
    $Z\gamma\gamma+W^{\pm}\gamma\gamma$       &    $8.84\pm 2.90$               \\ 
    $Z h + W^{\pm} h$                         &    $6.91\pm 2.56$         \\ 
    $h$                                       &    $0\pm 0$                   \\ 
    $\gamma\gamma$                            &    $0 \pm 0 $                     \\ 
     Total backgrounds                        &    $15.75 \pm 3.9$     \\ \hline  
     Signal                                   &    $297.8\pm 15.4$    \\ \hline \hline
    \end{tabular}
 \caption{Signal and backgrounds for $\gamma\gamma$+MET channel are shown with ${\cal L} = 300~\text{fb}^{-1}$
   at $\sqrt{s} = 14$ TeV, after the event selections and cuts discussed in the text are applied. 
   The signal events are for normalized cross section $\sigma = 5$ fb, and for $m_{\text{DM}} =$ 40 GeV
    and $m_{\rho} =$ 100 GeV.}
\label{MET-AA}
\end{table}

We find that the cut $\text{MET} > 150$ GeV on missing transverse energy together 
with the cut on $m_{\gamma\gamma}$ give the maximum sensitivity to the signal.
The former cut reduces background events with MET stemming from mismeasurement 
of identified physical objects like photons or soft radiations in the background
processes denoted by AA or H.
We show in Table.~\ref{MET-AA} the expected signal and background events at $\sqrt{s} = 14$ TeV with 
${\cal L} = 300~\text{fb}^{-1}$ for a signal benchmark with $m_{\text{DM}} = 40$ GeV and $m_{\rho} = 100$ GeV.  
According to the results given in Table.~\ref{MET-AA} backgrounds coming 
from $Z/W\gamma\gamma$ and $Z/Wh$ are the dominant ones. 

\subsection{Two b-jet channel}

Even though our signal with Higgs decaying into two b-quarks is associated with quite large SM backgrounds, 
this decay mode has the largest branching ratio, $Br(h \to b \bar b) = 0.577$. 
So it would be interesting to see if this channel can give a sensible sensitivity reach at the LHC.
In this study we simulate only the dominant backgrounds. We have ignored backgrounds with two bosons 
in the final state which have negligible contributions to the total background. 
By choosing a suitable cut for missing transverse energy, the QCD multi-jet backgrounds can be suppressed 
and we therefore do not take them into account in our simulations.   

We list here the most important backgrounds to the final state $b\bar b + \text{MET}$ as the following:

\begin{enumerate}
   \item $Z h$ with $h \to b\bar b$ and ($Z \to \nu \bar \nu$ or $Z \to \tau^+ \tau^-$ where $\tau$ 
      decay produces some missing energy). 
    This is an irreducible background and is denoted by CH in plots.
   \item $W h$ with $h \to b \bar b$ and $W \to l \nu$, denoted by CH.
   \item $Z b\bar b$ with $Z \to \nu \nu$ or $Z \to \tau^+ \tau^-$ (where $\tau$ 
   decay produces some missing energy), denoted by Ebb.
   \item $W b \bar b$ with $W \to l \nu$, denoted by Ebb.
   \item Higgs production  with $h \to b \bar b$, denoted by H.
   \item $t\bar t$ with $t \to b l^{+} \nu_l$ and $\bar t \to \bar b l^{-} \bar \nu_{l}$, 
         plus events in which one of the top quarks decay hadronically, denoted by TT. 
\end{enumerate}
\begin{figure}
\begin{minipage}{0.5\textwidth}
\includegraphics[width=\textwidth,angle =0]{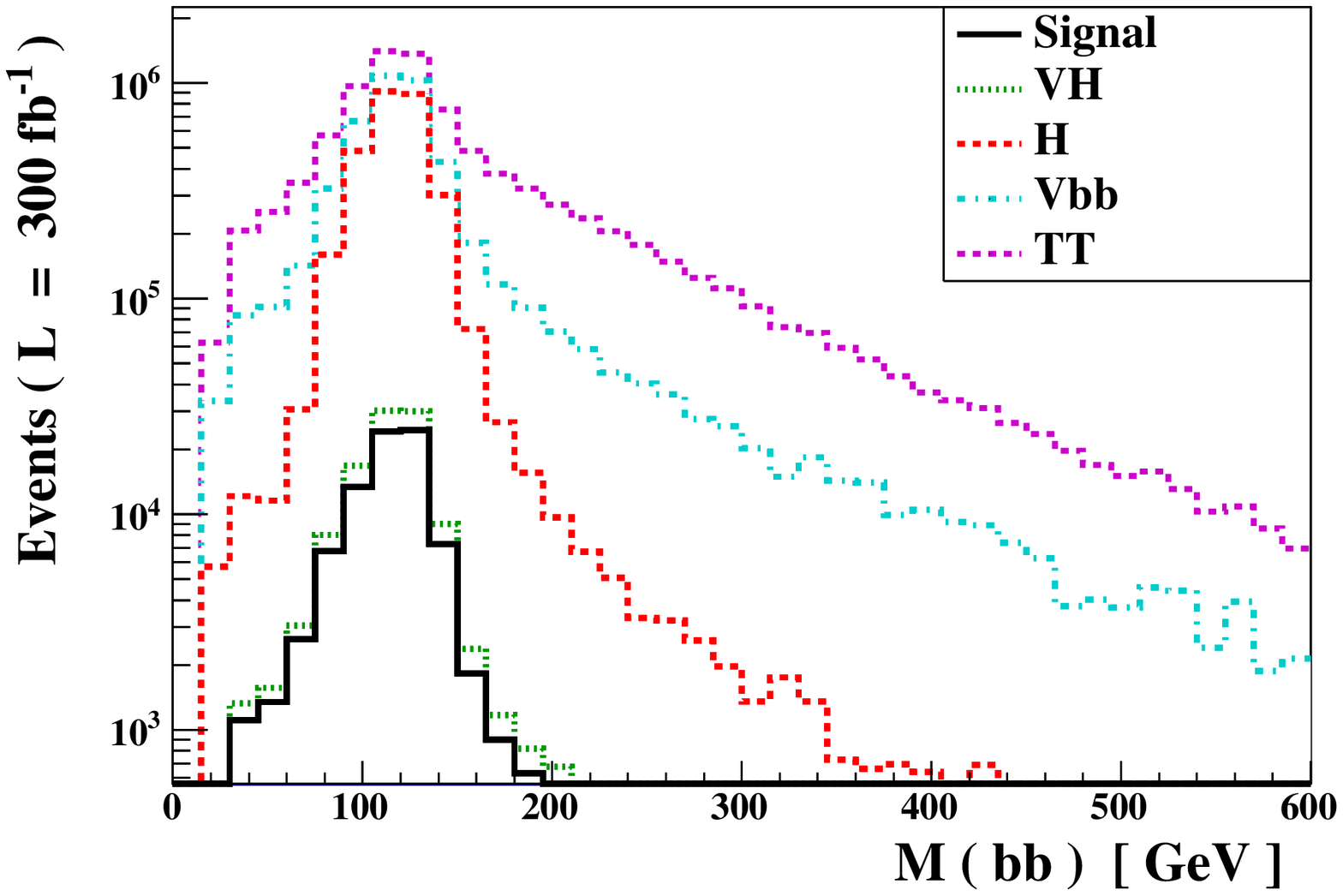}
\end{minipage}
\hspace{.1cm}
\begin{minipage}{0.5\textwidth}
\includegraphics[width=\textwidth,angle =0]{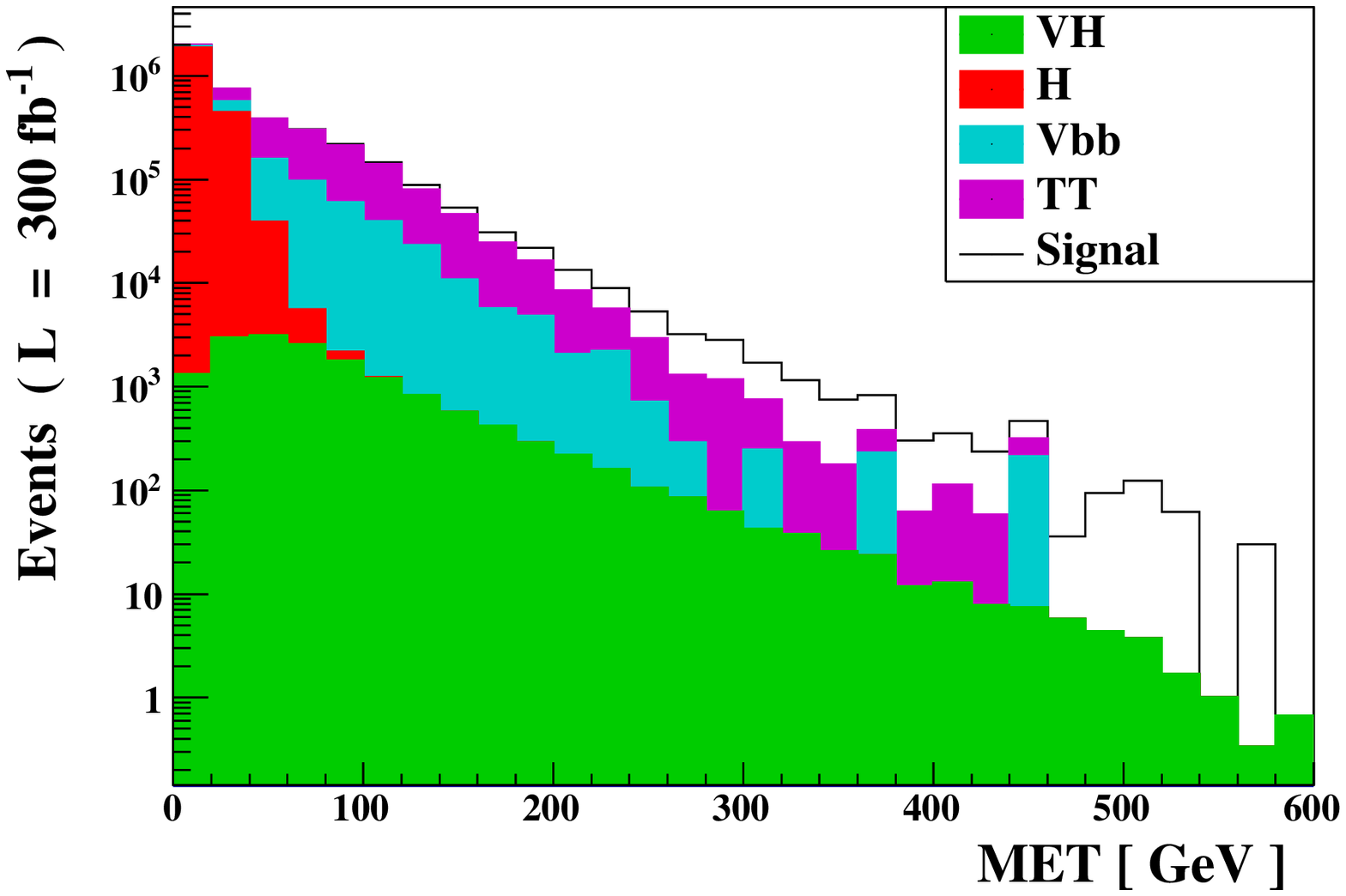}
\end{minipage}
\caption{
For $b\bar b+\text{MET}$ channel at $\sqrt{s} = 14$ TeV 
with $\cal{L} = $ $300~\text{fb}^{-1}$, in the left panel 
is the $b\bar b$ invariant mass for the signal and background processes  
and in the right panel is distributions of the missing transverse energy for the 
signal and backgrounds after applying the event selections discussed in the text. 
Signal is normalized to a nominal cross section of $1$ pb.}
\label{bb-invariant}
\end{figure}
Only events with two b-tagged jets with $p_T > 50$ will be kept in our event selection.   
Concerning charged leptons we select to veto those with transverse momentum 
$p_T > 20$ and rapidity $|\mu| < 2.5$. The signal region is defined with the invariant 
mass of b-quarks satisfying the cut $90~\text{GeV} < m_{bb} < 140~\text{GeV}$. This cut will 
reduce significantly backgrounds with two non-resonant b-quarks, like in top pair production case, $t\bar t$.  
Figure \ref{bb-invariant} shows $b\bar b$ invariant mass distribution and, signal and background events 
with signal normalized to $1$ pb at $\sqrt{s} = 14$ TeV and ${\cal L} = 300~\text{fb}^{-1}$.
We find that selecting a cut as $\text{MET} > 320$ GeV will optimize the signal to background ratio. 
As we said earlier, background cross sections are calculated at leading order but they are scaled so as to 
incorporate higher order corrections. 

\begin{table}
\centering
\begin{tabular}{p{4.5cm} p{3cm}}
 \hline \hline
    $pp \to b {\bar b} + \slashed{E}_{T}$    &    $\slashed{E}_{T} > 320$ GeV \\ \hline \hline 
    $Z b{\bar b}+W^{\pm}b{\bar b}$           &    $411.4 \pm 20.3$               \\ 
    $Z h + W^{\pm} h$                        &    $146 \pm 12.1 $         \\ 
    $h$                                      &    $0 \pm 0$                   \\ 
    $t{\bar t}$                              &    $860.6 \pm 29.3$                  \\ 
     Total backgrounds                       &    $1418.1 \pm 37.7 $     \\ \hline  
     Signal                                  &    $3120 \pm 55.6 $    \\ \hline \hline
    \end{tabular}
 \caption{Signal and backgrounds for $b \bar b$+MET channel are shown with ${\cal L} = 300~\text{fb}^{-1}$
   at $\sqrt{s} = 14$ TeV, after the event selections and cuts discussed in the text are applied. 
   The signal events are for normalized cross section $\sigma = 1$ pb, and for $m_{\text{DM}} =$ 40 GeV
    and $m_{\rho} =$ 100 GeV.}
\label{MET-BB}
\end{table}
The cross section for $Z/W +b\bar b$ is corrected by a factor $k= 1.48$ \cite{Cordero:2009kv}.
For $Z/W +h$ cross section and $h$ cross section, $k= 1.18$ \cite{Alwall:2014hca} and $k= 2$ \cite{Alwall:2014hca} 
are used respectively. The LO cross section for $t\bar t$ production is corrected by $k = 1.47$ \cite{Alwall:2014hca}.    
In Table.~\ref{MET-BB} we show signal and background events for $h\to b\bar b$ channel for 
the event selections and cuts described above. In this channel even after the cuts, the remaining backgrounds are sizable.

\subsection{Selection efficiencies and exclusions}
In this section we present our main results for the efficiencies, upper bounds on the signal 
cross section and upper bounds on the Yukawa coupling $g_{\chi}$ (coupling between DM and SM Higgs) 
for two channels $h \to \gamma\gamma$ and $h \to b\bar b$ based on the cuts discussed above. 
To find the $95\%$ CL exclusion for 
our signal benchmark points neglecting systematic uncertainties, we define significance as 
$\mathbb{S} = S/\sqrt{S+B}$, for signal events $S$ and total background $B$. 
When no signal is observed, the upper limit on the cross section (or excluded cross section) 
for a given signal benchmark point is obtained by requiring a significant of $\sim 2\sigma$.      

\begin{figure}
\begin{minipage}{0.51\textwidth}
\includegraphics[width=\textwidth,angle =0]{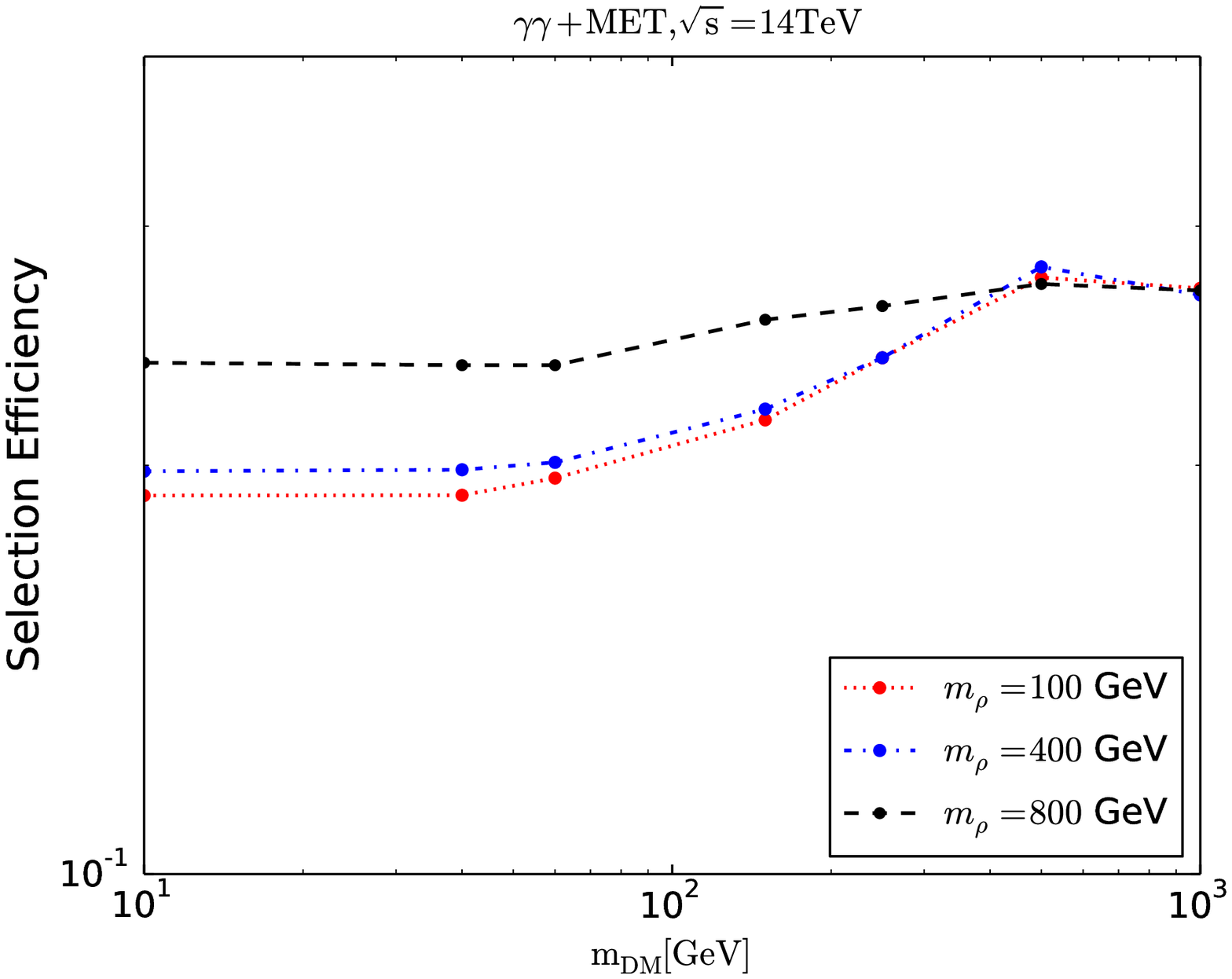}
\end{minipage}
\hspace{.1cm}
\begin{minipage}{0.51\textwidth}
\includegraphics[width=\textwidth,angle =0]{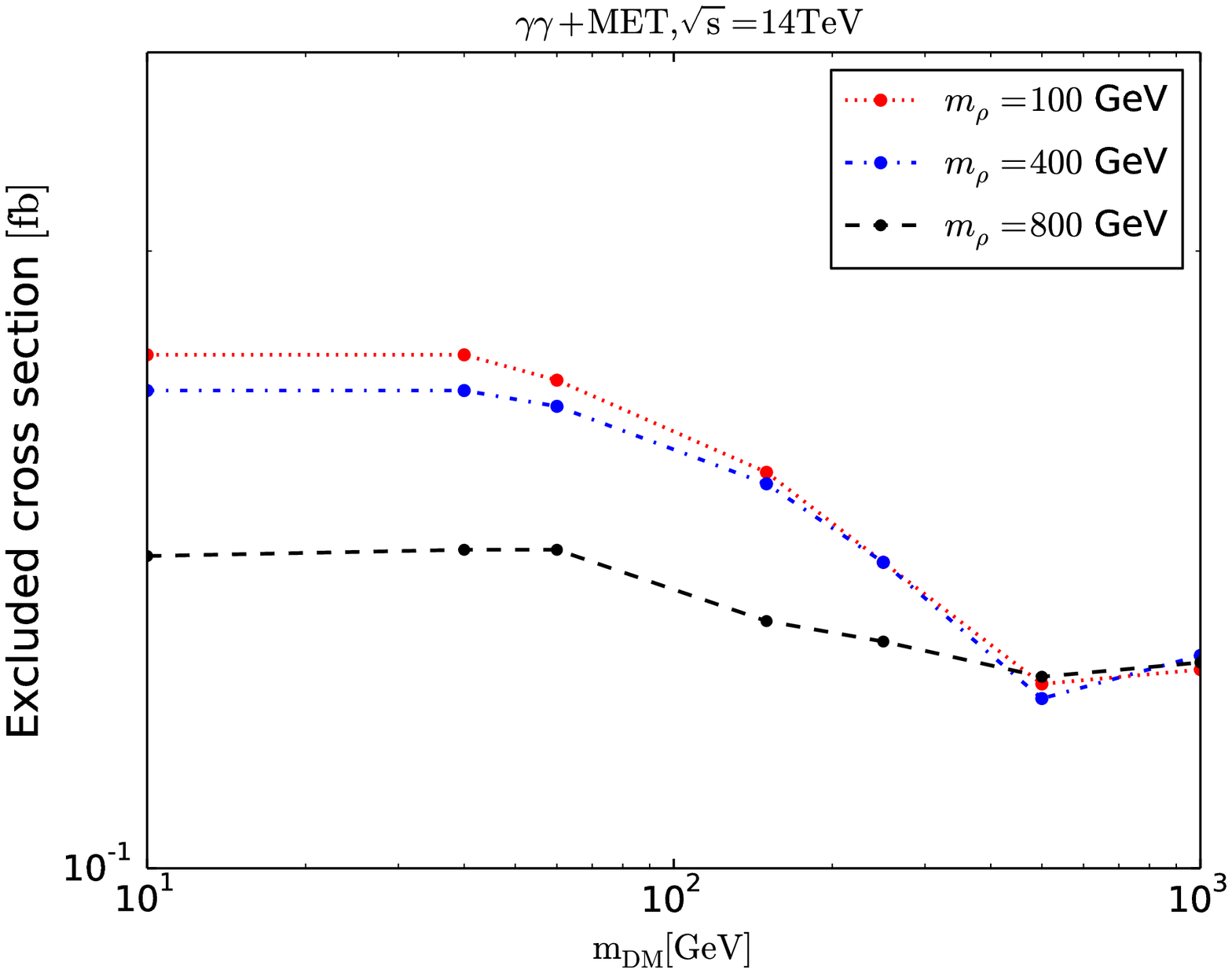}
\end{minipage}
\caption{Shown are the selection efficiency in the $\gamma\gamma$ + MET channel (left panel) and 
   upper limit on the cross section $\sigma (p p \to \gamma\gamma \chi \bar \chi)$ 
   for $\sqrt{s} = 14$ TeV and ${\cal L} = 300~\text{fb}^{-1}$ at the LHC.}
\label{aa-efficiency-upper}
\end{figure}

\begin{figure}
\begin{minipage}{0.51\textwidth}
\includegraphics[width=\textwidth,angle =0]{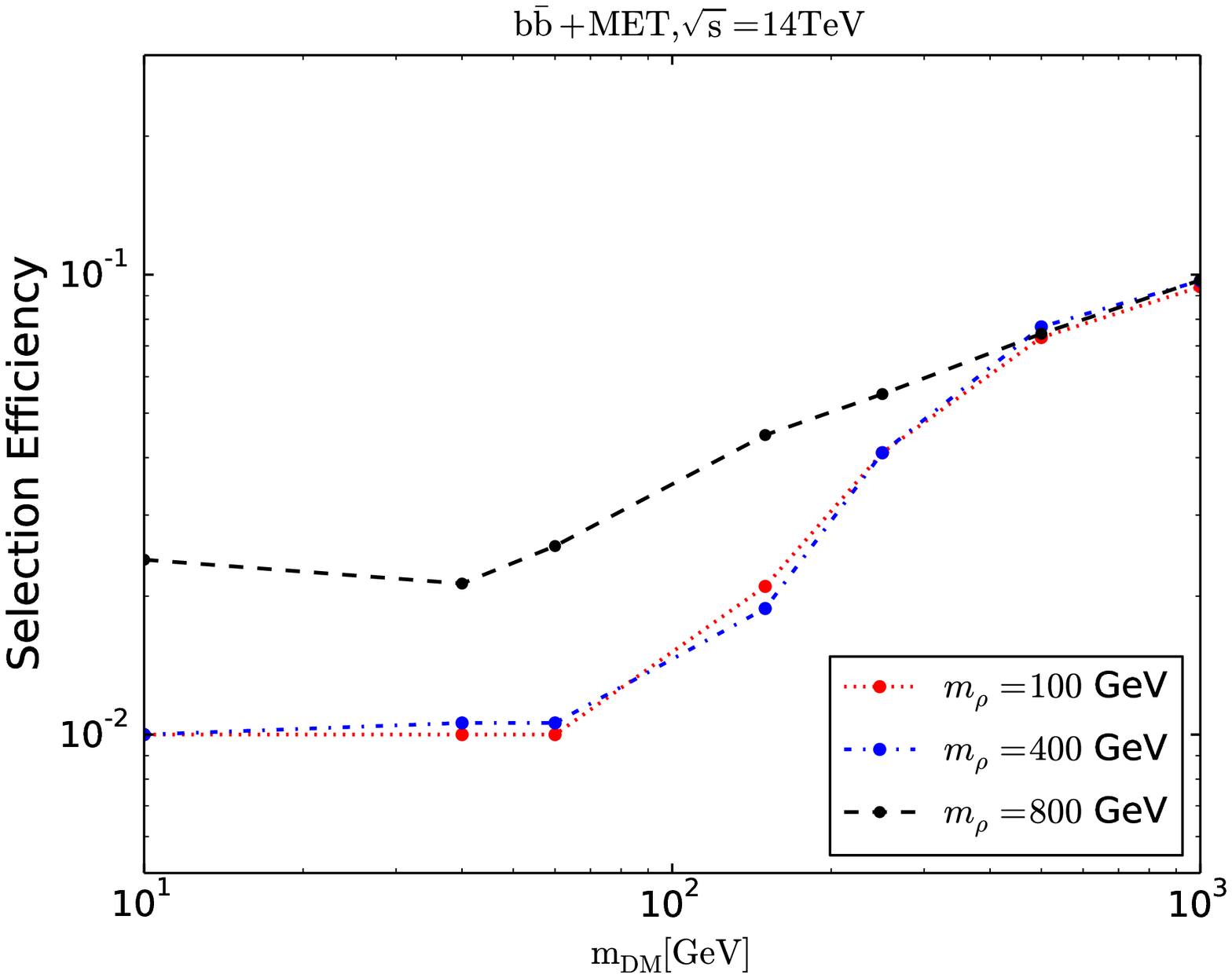}
\end{minipage}
\hspace{.1cm}
\begin{minipage}{0.51\textwidth}
\includegraphics[width=\textwidth,angle =0]{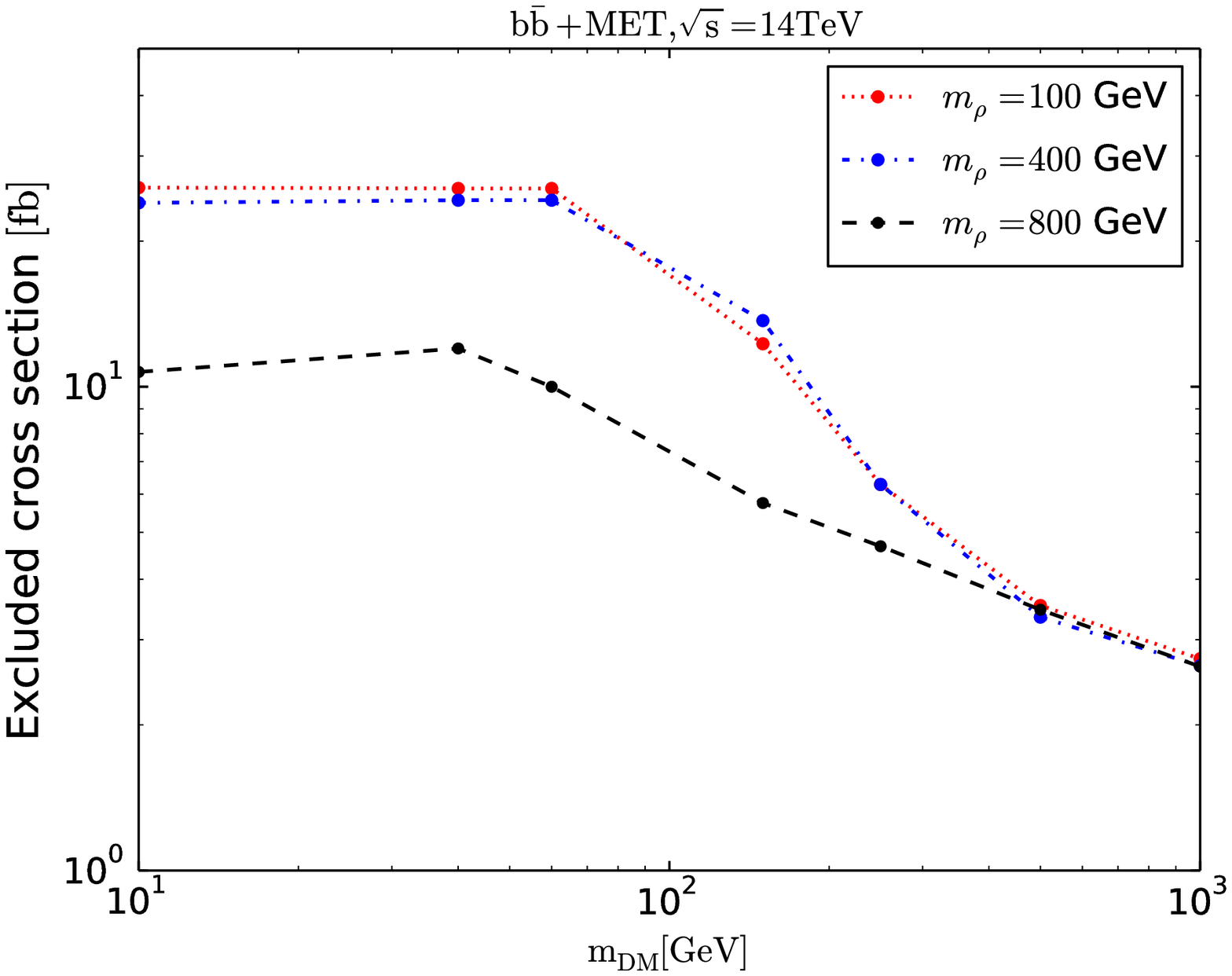}
\end{minipage}
\caption{Shown are the selection efficiency in the $b\bar b$ + MET channel (left panel) and 
   upper limit on the cross section $\sigma (p p \to b\bar b \chi \bar \chi)$ 
   for $\sqrt{s} = 14$ TeV and ${\cal L} = 300~\text{fb}^{-1}$ at the LHC.}
\label{bb-efficiency-upper}
\end{figure}

We first obtain the selection efficiencies as a function of DM mass for various benchmark points and 
then we evaluate the upper bound on the cross section respecting the criterion $\mathbb{S} \sim 2\sigma$.
Our result for the channel with $\gamma\gamma+\text{MET}$ in the final state is shown 
in Fig.~\ref{aa-efficiency-upper} for the selection efficiency, $\epsilon_{s}$, (left panel) and for 
the excluded cross section, $\sigma_{s}$, (right panel) at $\sqrt{s} = 14$ TeV and ${\cal L} = 300~\text{fb}^{-1}$.
The excluded cross section is obtained by setting $S = \sigma_{s} \epsilon_{s} {\cal L}$ in 
the relation for the significance $\mathbb{S}$ and solving the equation for $\sigma_{s}$. 
It can be seen from the solution for $\sigma_{s}$ that the excluded cross section decreases 
by increasing the selection efficiency.  
Figure \ref{bb-efficiency-upper} shows results for the $b\bar b+\text{MET}$ channel. 
In both channels, as expected, better efficiency is achieved for larger DM mass 
independent of the value for $m_\rho$. 
Moreover, the efficiency gets improved for larger $m_{\rho}$.
We also see that due to larger backgrounds, upper bound on the 
cross section $\sigma(p p \to b\bar b \chi \chi)$ 
is weaker than the one on the cross section $\sigma(p p \to \gamma\gamma \chi \chi)$.

Next, we move on to evaluate the upper limit on the Yukawa coupling. 
We obtain the upper limit for fixed mixing angle at $\sin \beta = 0.1$. 
We remind that the signal production cross section is proportional to $g_{\chi}^2$ 
so that the results in Fig~\ref{sig-production} along with the upper bounds on  
$\sigma(p p \to \gamma \gamma \chi \chi)$  and $\sigma(p p \to b\bar b \chi \chi)$ 
provided by Fig.~\ref{aa-efficiency-upper} and Fig.~\ref{bb-efficiency-upper} 
can be used to achieve the upper limit on $g_{\chi}$ for different benchmark points. 

\begin{figure}
\begin{minipage}{0.51\textwidth}
\includegraphics[width=\textwidth,angle =0]{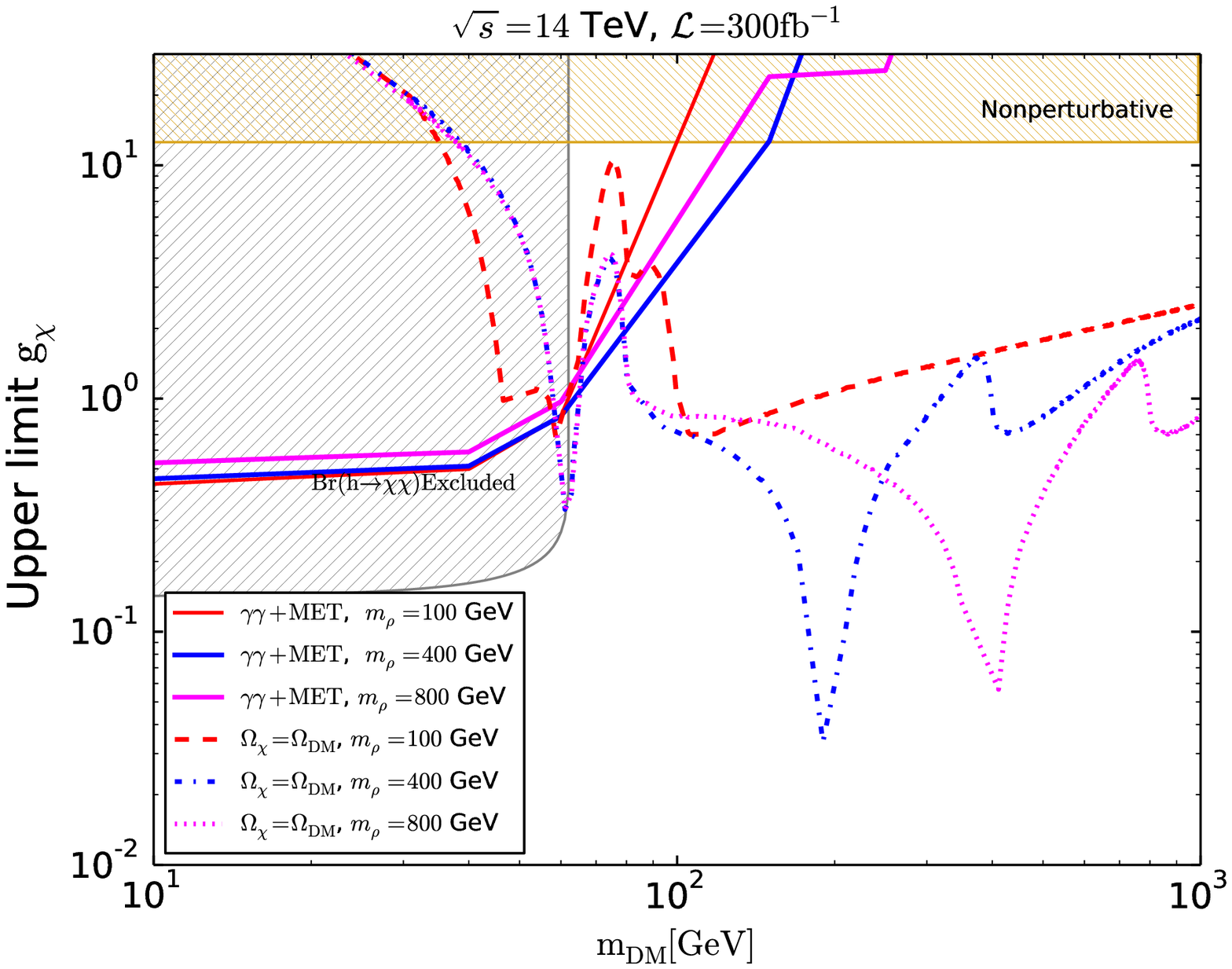}
\end{minipage}
\hspace{.01cm}
\begin{minipage}{0.51\textwidth}
\includegraphics[width=\textwidth,angle =0]{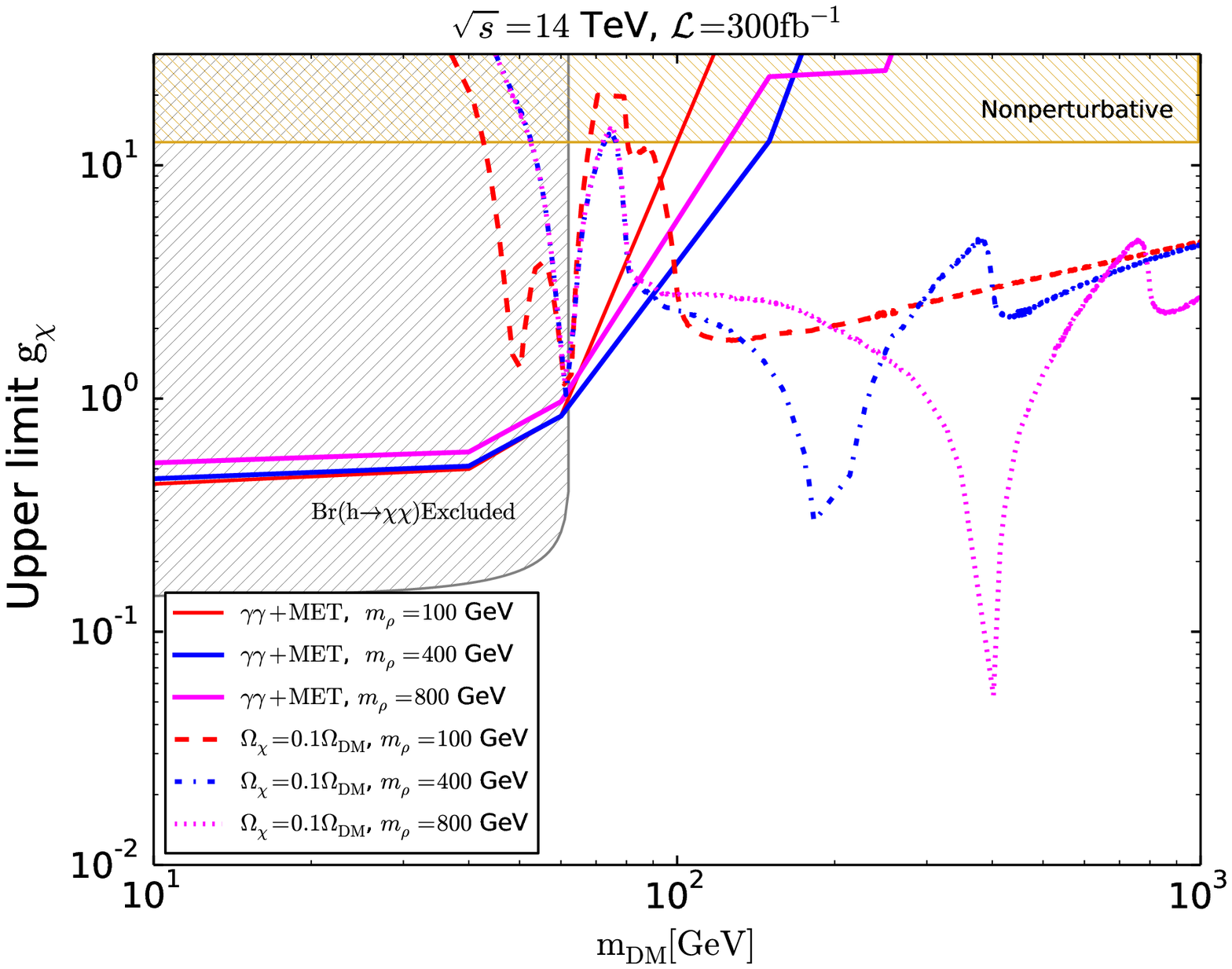}
\end{minipage}

\begin{minipage}{0.51\textwidth}
\includegraphics[width=\textwidth,angle =0]{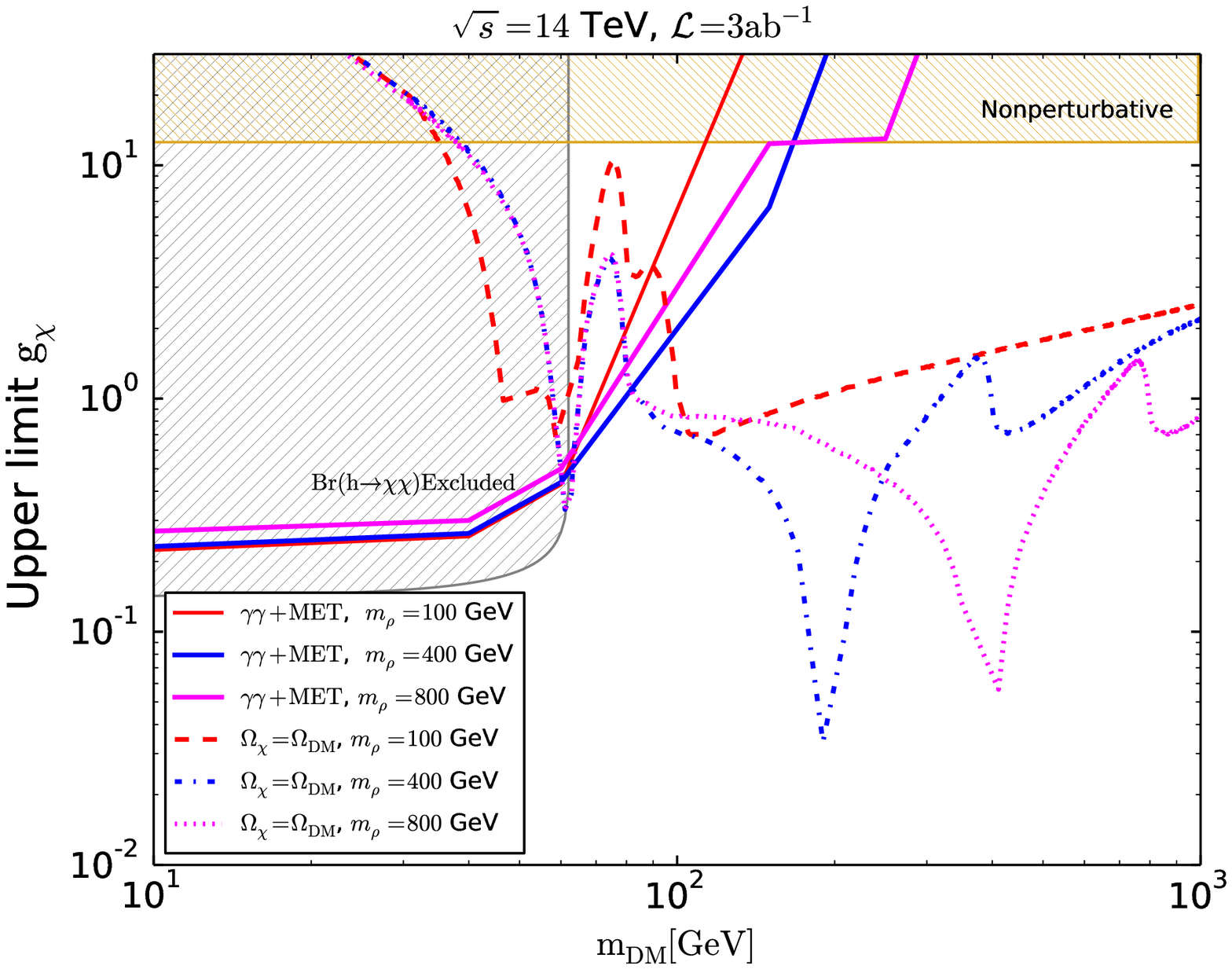}
\end{minipage}
\hspace{.01cm}
\begin{minipage}{0.51\textwidth}
\includegraphics[width=\textwidth,angle =0]{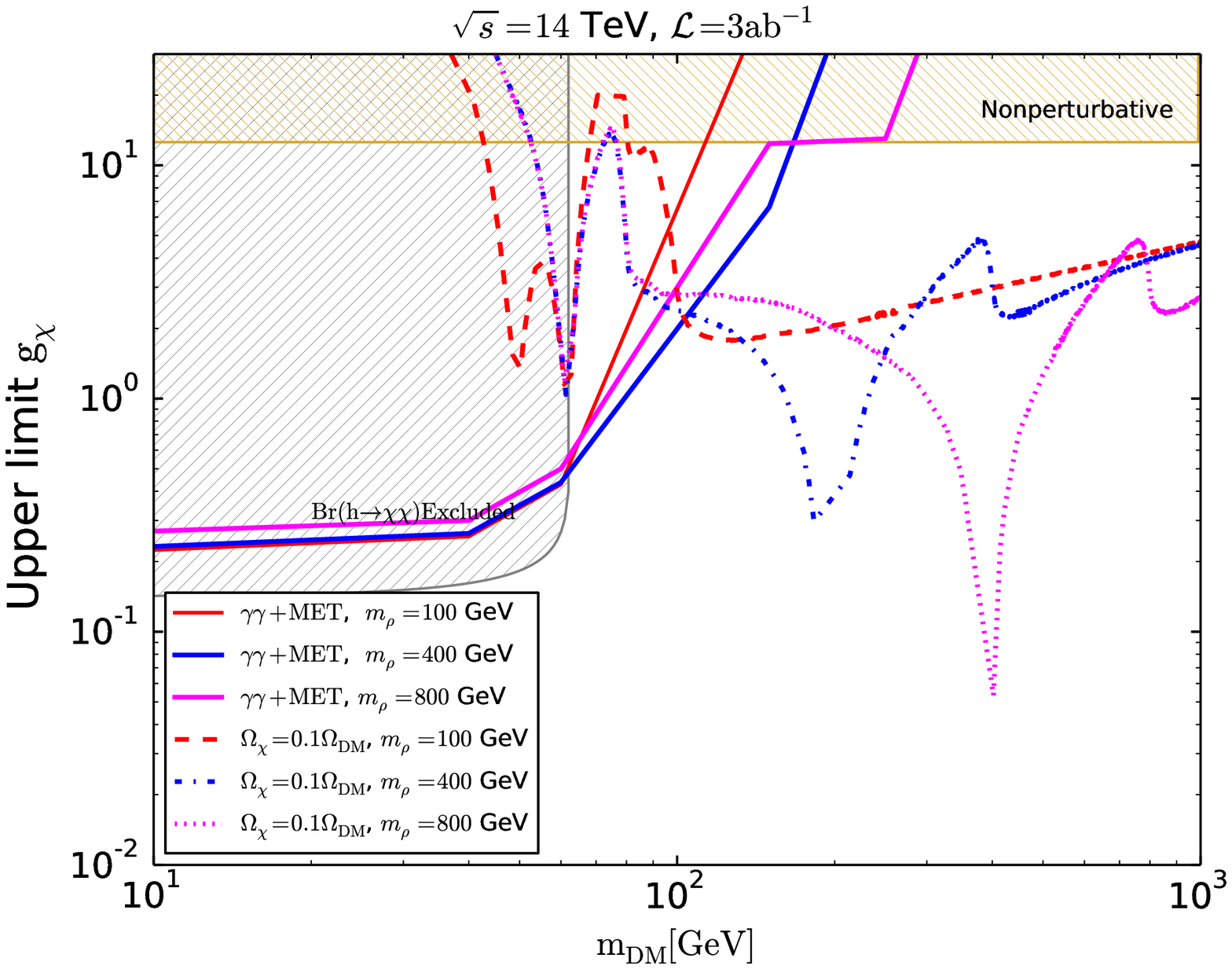}
\end{minipage}

\caption{Shown are the projected LHC mono-Higgs sensitivities at $\sqrt{s} = 14$ TeV 
in the $\gamma \gamma+\text{MET}$ final states with ${\cal L} = 300~\text{fb}^{-1}$ 
for plots on the top and with ${\cal L} = 3~\text{ab}^{-1}$ for plots 
on the bottom. All solid lines are contours (corresponding to $95\%$ CL upper limit) 
which exclude larger coupling, and 
broken lines show viable points in the $m_{\chi}-g_{\chi}$ plane for various $m_{\rho}$.
Comparison made between plots on the right and plots on the left for two cases, 
when $\Omega_{\chi} = \Omega_{\text{DM}}$ or $\Omega_{\chi} = 0.1 \Omega_{\text{DM}}$. 
The horizontal shaded area is excluded because it violates perturbativity.
The shaded region on the left is excluded by the invisible Higgs decay width measurements.}
\label{AA-upper-g}
\end{figure}

Given our projected mono-Higgs sensitivities, in Fig.~\ref{AA-upper-g} we show contours 
which indicate the 14 TeV LHC upper bounds on the Yukawa coupling, $g_{\chi}$, from 
mono-Higgs searches with $\gamma\gamma+\text{MET}$ final states. The contours are obtained
for two prospective integrated luminosities 
${\cal L} = 300~\text{fb}^{-1}$ and ${\cal L} = 3~\text{ab}^{-1}$ as a function of DM mass and the coupling.
In addition, for each benchmark point we show the viable 
region in the $m_{\chi}-g_{\chi}$ plane respecting the observed DM relic density 
($\Omega_{\chi} = \Omega_{\text{DM}}$) and also the viable regions 
with $\Omega_{\chi} = 0.1 \Omega_{\text{DM}}$ for three pseudoscalar masses $m_{\rho} = 100, 400$ and $800$ GeV.
It is evident in the plots that the viable value for $g_{\chi}$ drops at $m_{\text{DM}} \sim m_{\rho}/2$ 
and $m_{\text{DM}} \sim m_{h}/2$. The reason hinges in the fact the annihilation cross section peaks when 
the mediator mass is about twice the DM mass which is called the resonance region. Since the annihilation 
cross section is proportional to $g_{\chi}^2$, in order to get the observed relic 
density a smaller value for $g_{\chi}^2$ is picked up.

In the plane $m_{\chi}-g_{\chi}$, for $m_{\text{DM}} < m_h/2$ there is already strong constraint from invisible Higgs decay 
measurements which is slightly stronger than the upper limits from the mono-Higgs searches
for the integrated luminosity ${\cal L} = 300~\text{fb}^{-1}$. However, at the larger luminosity 
the bounds from the two constraints are comparable. 

For $m_{\text{DM}} > m_h/2$, the mono-Higgs sensitivities are higher for the intermediate mediator mass, $m_{\rho} = 400$ GeV. 
For the case with $\Omega_{\chi} = 0.1 \Omega_{\text{DM}}$ and $m_{\rho} = 400$ GeV, we find that the respective contour 
excludes DM masses smaller than $\sim 90$ GeV and $\sim 100$ GeV at 
${\cal L} = 300~\text{fb}^{-1}$ and ${\cal L} = 3~\text{ab}^{-1}$ respectively. 
We note that perturbativity condition excludes regions with $g_{\chi} > 4\pi$.

\begin{figure}
\begin{minipage}{0.51\textwidth}
\includegraphics[width=\textwidth,angle =0]{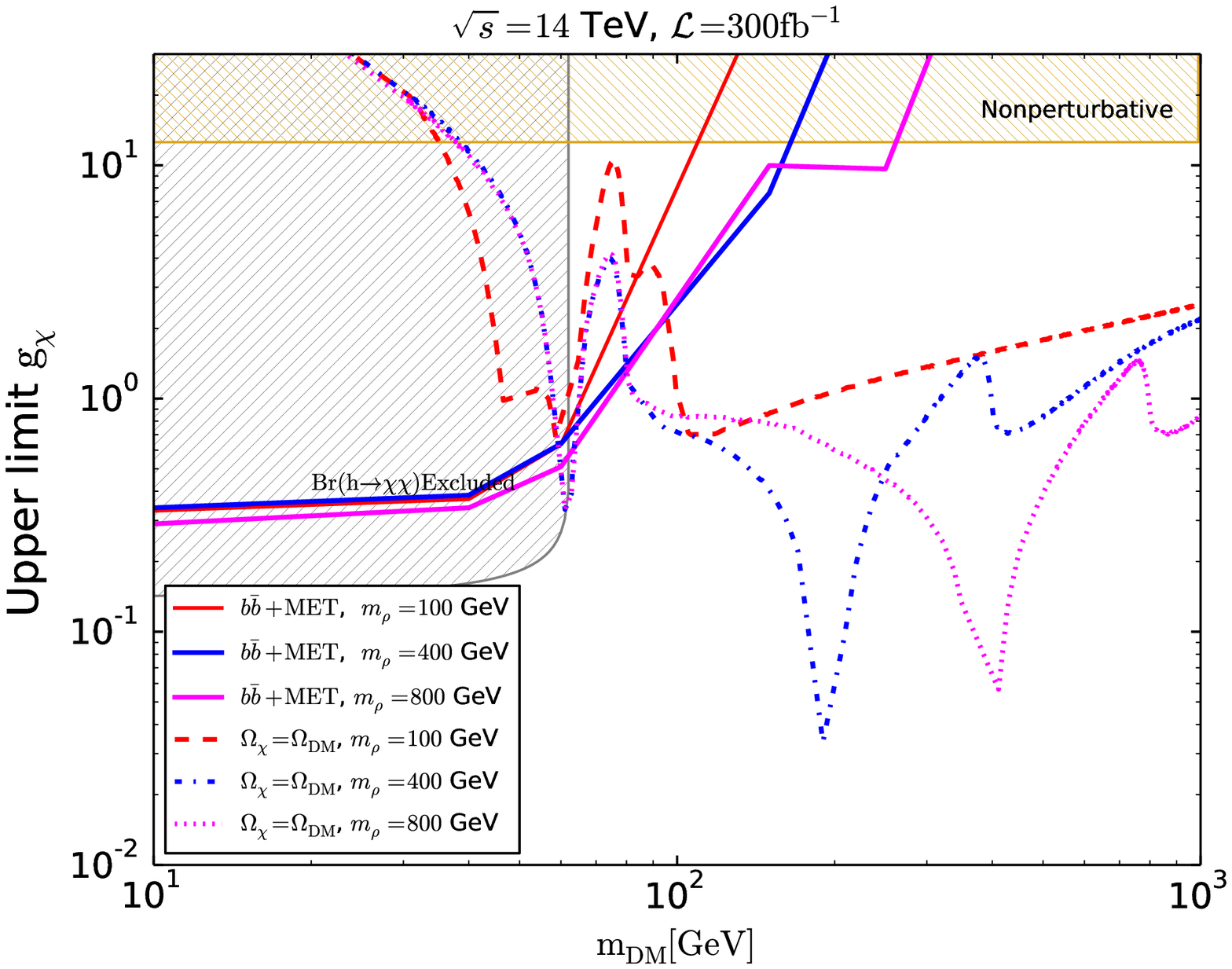}
\end{minipage}
\hspace{.01cm}
\begin{minipage}{0.51\textwidth}
\includegraphics[width=\textwidth,angle =0]{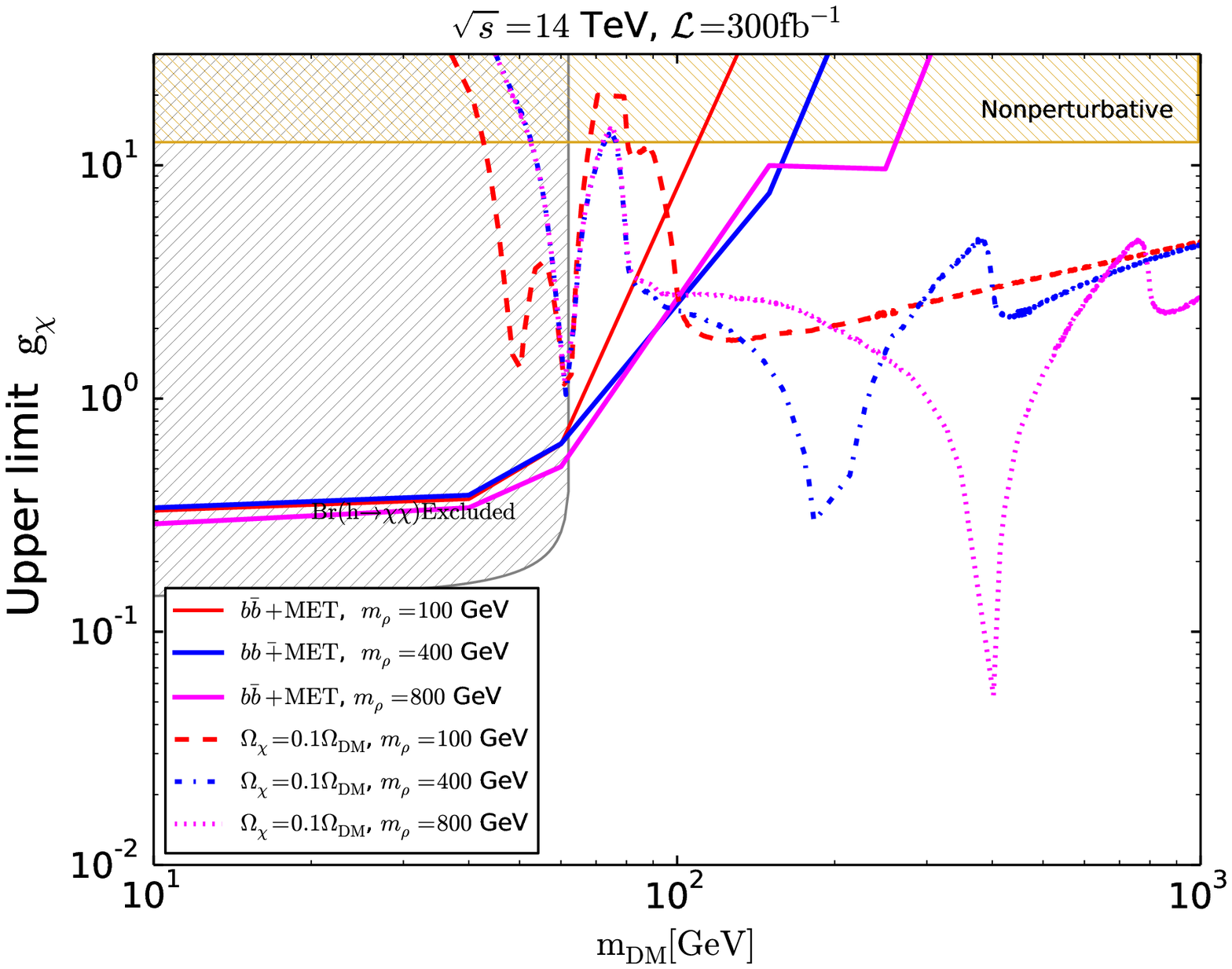}
\end{minipage}

\begin{minipage}{0.51\textwidth}
\includegraphics[width=\textwidth,angle =0]{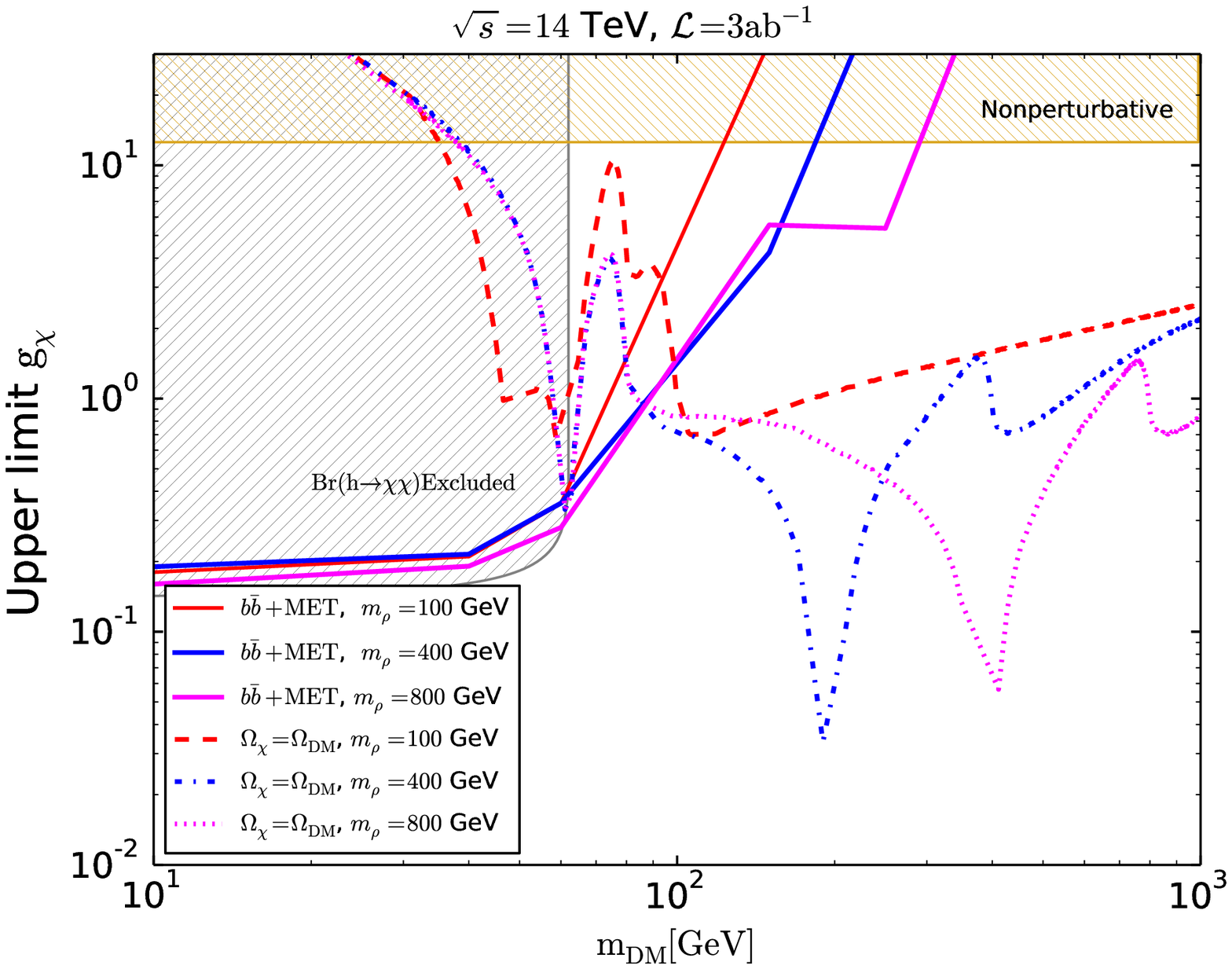}
\end{minipage}
\hspace{.01cm}
\begin{minipage}{0.51\textwidth}
\includegraphics[width=\textwidth,angle =0]{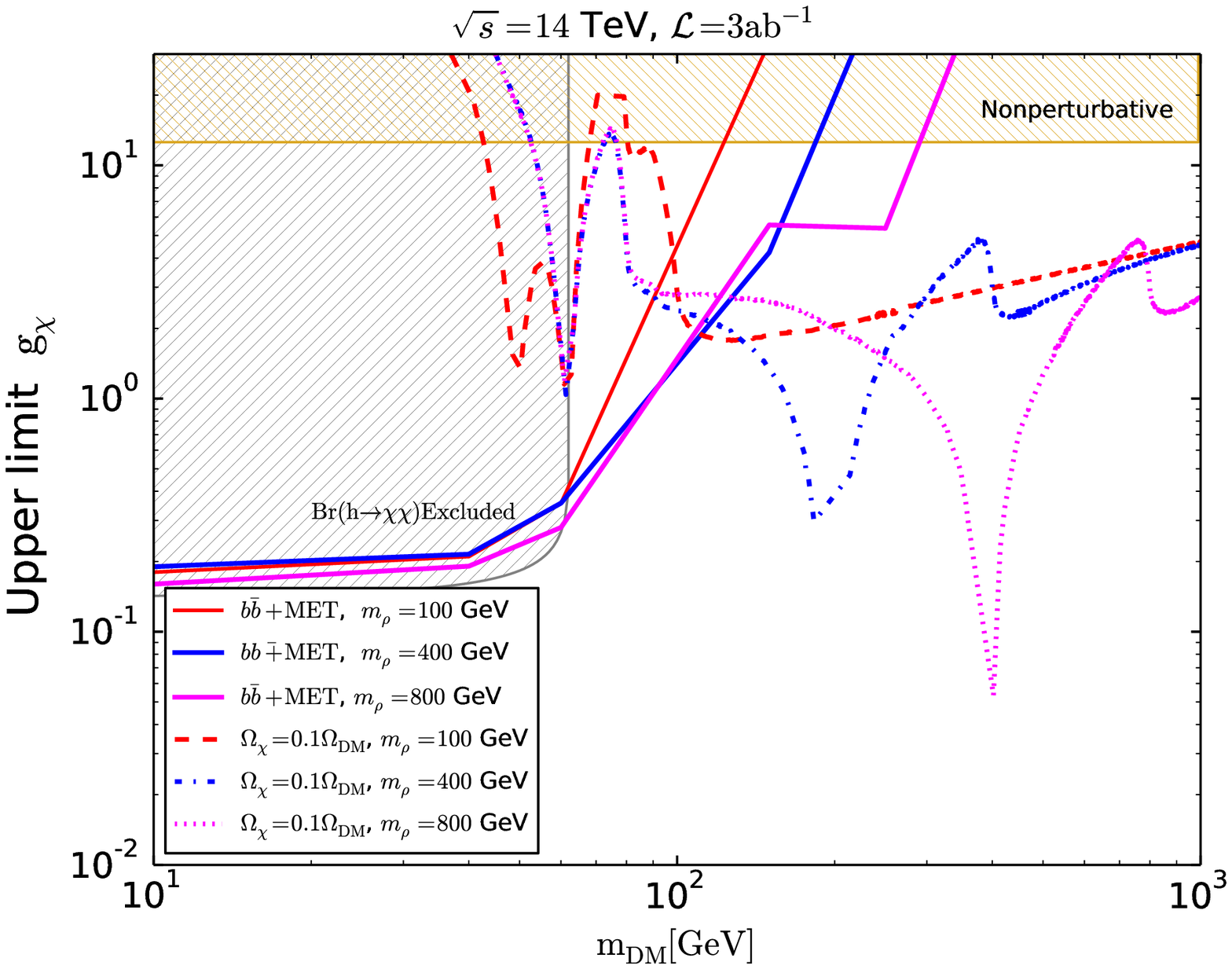}
\end{minipage}

\caption{Shown are the projected LHC mono-Higgs sensitivities at $\sqrt{s} = 14$ TeV 
in the $b \bar b +\text{MET}$ final states with ${\cal L} = 300~\text{fb}^{-1}$ 
for plots on the top and with ${\cal L} = 3~\text{ab}^{-1}$ for plots 
on the bottom. All solid lines are contours (corresponding to $95\%$ CL upper limit) 
which exclude larger coupling, and 
broken lines show viable points in the $m_{\chi}-g_{\chi}$ plane for various $m_{\rho}$.
Comparison made between plots on the right and plots on the left for two cases, 
when $\Omega_{\chi} = \Omega_{\text{DM}}$ or $\Omega_{\chi} = 0.1 \Omega_{\text{DM}}$. 
The horizontal shaded area is excluded because it violates perturbativity.
The shaded region on the left is excluded by the invisible Higgs decay width measurements.}
\label{BB-upper-g}
\end{figure}

We then continue our analysis for the $b\bar b + \text{MET}$ channel. 
Our results are shown for two integrated luminosities in Fig.~\ref{BB-upper-g} when 
fermionic DM constitutes fully the observed DM relic density or when it 
only makes up $10\%$ of the observed relic density. 
The upshot is that in this model, the mono-Higgs searches with $b\bar b + \text{MET}$ in the final state 
has a stronger exclusion power compared with the $\gamma\gamma + \text{MET}$ channel for the Yukawa coupling. 
It is clearly seen in Fig.~\ref{BB-upper-g} that the mono-Higgs constraints get stronger with the mass 
of the mediator, $m_{\rho}$. The strongest constraint belongs to the case in which 
$\Omega_{\chi} = 0.1 \Omega_{\text{M}}$, ${\cal L} = 3~\text{ab}^{-1}$ and $m_{\rho} = 800$ GeV, 
where regions with $m_{\text{DM}} \lesssim 120$ GeV are excluded.

There are other LHC searches that can potentially constrain the model parameter space, e.g., the pseudoscalar mass. 
In these searches, upper limits are found for the cross section in the processes 
$p p \to \rho \to (W^+ W^- , ZZ)$ \cite{ATLAS:2016oum,ATLAS:2016npe,ATLAS:2016kjy} and
   $p p \to \rho \to hh$ \cite{ATLAS:2016ixk,CMS:2016knm}. In our model, vector boson production cross
 section is suppressed by a factor $\sin^4 \beta$ and the di-Higgs production cross section is 
 suppressed by a factor $\sin^2 \beta$. Therefore one expects that the cross sections reside 
 below the current upper limits for a wide range of the pseudoscalar mass. 
 Recently, in a study \cite{Baek:2017vzd} within the same fermionic DM model, 
 it is confirmed that these searches put no constraints on the pseudoscalar mass.
 In addition, these searches cannot constrain the coupling $g_{\chi}$, 
   since in the cross section $\sigma (p p \to \text{diboson}) 
   \sim \sigma (p p \to \rho) \times Br(\rho \to \text{diboson})$, the coupling $g_{\chi}$ 
   appears only in numerator of $Br$ and plays an insignificant role.

\section{Conclusions}
\label{conclusion}
We considered a model with a fermionic DM candidate which interacts with the SM Higgs 
due to a mixing between a pseudoscalar singlet field and the Higgs field. 
The specific nature of the DM-SM Yukawa interaction leads to a velocity suppressed elastic 
scattering cross section of DM-nucleon. Therefore near future direct detection experiments 
would not be able to probe the Yukawa interaction. 
In the model that assumes DM and the Higgs to have direct interaction,
it seems plausible to search for DM signature at a collider like the LHC 
which is a Higgs production machine.  

We have studied in this work, processes with missing energy coming from DM production 
associated with a Higgs boson. Our analysis are done for two channels with $\gamma\gamma + \text{MET}$
and $b \bar b + \text{MET}$ in the final states. 
We simulated the SM background processes and signal events for various benchmark points in both channels
at $\sqrt{s} = 14$ TeV with ${\cal L} = 300~\text{fb}^{-1}$.
After event selections, backgrounds for the $\gamma\gamma+\text{MET}$ channel are found to be quite small 
while this is not the case for $b \bar b + \text{MET}$  channel. 

The selection efficiencies, upper bounds on the signal cross section and upper limits on the Yukawa
coupling, $g_{\chi}$, are estimated for both channels. Our main results along with constraints from 
invisible Higgs decay width measurements, observed DM relic density, perturbativity requirement and
electroweak measurements are collected in Fig.~\ref{AA-upper-g} and Fig.~\ref{BB-upper-g}.  

It turned out that the $b \bar b+\text{MET}$ channel, having the largest branching ratio, gives better
LHC sensitivity reach and therefore constrains the Yukawa coupling more strongly.

Finally, we point out that there are regions in the model parameter space 
compatible with all the constraints mentioned above which can be reached by future LHC searches.

\section{Acknowledgments}
We are very grateful to Lorenzo Basso for helpful discussions.
We would like to thank Hossein Ghorbani which was involved in the early stages of this research.
\label{Ack}


\bibliography{ref}
\bibliographystyle{utphys}

\end{document}